\documentclass[aps,prb,longbibliography,amsmath,twocolumn,amssymb,nofootinbib,superscriptaddress,showpacs,notitlepage]{revtex4-2}

\usepackage[T1]{fontenc}
\usepackage[english]{babel}
\usepackage{color}
\usepackage{graphicx}
\usepackage{epsfig}
\usepackage{epstopdf}
\usepackage{url}
\usepackage[normalem]{ulem}
\usepackage{amsthm}
\usepackage{amsmath}
\usepackage{epigraph}

\usepackage{float}
\makeatletter
\let\newfloat\newfloat@ltx
\makeatother
\usepackage{algorithm}
\usepackage[noend]{algpseudocode}
\usepackage{algorithmicx}

\usepackage[dvipsnames]{xcolor}

\newcommand*\rfrac[2]{{}^{#1}\!/_{#2}}

\def\be{\begin{equation}}
\def\ee{\end{equation}}
\def\bea{\begin{eqnarray}}
\def\eea{\end{eqnarray}}
\def\bi{\begin{itemize}}
\def\ei{\end{itemize}}
\def\bin{\begin{enumerate}}
\def\ein{\end{enumerate}}

\begin{document}

\title{Population annealing with topological defect driven nonlocal updates
for spin systems with quenched disorder}

%%%%%%%%%%%%%%%%%%%%%%%%%%%%%%%%%%%%%%%%%%%%%%%%%%%%%%%%%%%%%%%%%%%%%%%%%%%%%%%
\author{David Cirauqui}
\affiliation{Quside Technologies SL, Mediterranean Technology Park, 08860 Castelldefels (Barcelona), Spain}
\affiliation{ICFO - Institut de Ciències Fot\`oniques, The Barcelona Institute of Science and Technology, 08860 Castelldefels (Barcelona), Spain}

\author{Miguel \'Angel Garc\'ia-March}
\affiliation{Instituto Universitario de Matem\'atica Pura y Aplicada, Universitat Polit\`ecnica de
Val\`encia, 46022 Val\`encia, Spain}

\author{J. R. M. Saavedra}
\affiliation{Quside Technologies SL, Mediterranean Technology Park, 08860 Castelldefels (Barcelona), Spain}

\author{Maciej Lewenstein}
\affiliation{ICFO - Institut de Ciències Fot\`oniques, The Barcelona Institute of Science and Technology, 08860 Castelldefels (Barcelona), Spain}
\affiliation{ICREA - Instituci\'o Catalana de Recerca i Estudis Avan\c cats, E-08010 Barcelona, Spain}

\author{Przemys{\l}aw R. Grzybowski}
\affiliation{ICFO - Institut de Ciències Fot\`oniques, The Barcelona Institute of Science and Technology, 08860 Castelldefels (Barcelona), Spain}
\affiliation{Institute of Spintronics and Quantum Information, Faculty of Physics, Adam Mickiewicz University, Umultowska 85, 61-614 Pozna{\'n}, Poland}

\date{\today}

\begin{abstract}

Population Annealing, one of the currently state-of-the-art algorithms for solving spin-glass systems, sometimes finds hard disorder instances for which its equilibration quality at each temperature step is severely damaged. In such cases one can therefore not be sure about having reached the true ground state without vastly increasing the computational resources. In this work we seek to overcome this problem by proposing a quantum-inspired modification of Population Annealing. 
Here we focus on three-dimensional random plaquette gauge model
which ground state energy problem seems to be much harder to solve than standard spin-glass Edwards-Anderson model. In analogy to the Toric Code, by allowing single bond flips we let the system explore non-physical states, effectively expanding the configurational space by the introduction of topological defects that are then annealed through an additional field parameter. The dynamics of these defects allow for the effective realization of non-local cluster moves, potentially easing the equilibration process. We study the performance of this new method in three-dimensional random plaquette gauge model lattices of various sizes and compare it against Population Annealing. With that we conclude that the newly introduced non-local moves are able to improve the equilibration of the lattices, in some cases being superior to a normal Population Annealing algorithm with a sixteen times higher computational resource investment.
\end{abstract}

\maketitle

\section{Introduction}

Spin-glass models and, more generally, spin systems with quenched disordered interactions, consist of a set of particles or nodes, each taking a value from a range of possibilities (binary in case of the Ising models \cite{brush}) with randomly valued interactions between them, where the topology of the connections can be arbitrarily complex. Due to spin frustration and randomness such models exhibit tremendously rich behaviors and have long been one of the main focuses of statistical physics. Analytically challenging even at the mean-field level \cite{parisi2007mean} and numerically hard for non-planar topologies \cite{Barahona1982}, %prove to be analytically challenging, even intractable; thus, 
their study continues to rely mostly on numerical simulations.

In the last decades, spin-glass models have proven useful beyond their applications in physics, for example as models related to the behavior of neural networks \cite{Amit1985, vanHemmen1986} and to the real-world optimization problems, especially for those that accept a Quadratic Unconstrained Binary Optimization (QUBO) mapping \cite{Lucas2014, Lucas2019}. In the QUBO picture, an optimization problem is mapped to a set of binary variables such that its ground state configuration encodes the optimal solution to the original problem. QUBO mapping applications range from fields as diverse as protein folding \cite{Irback2022, Oliveira2018}, to logistics \cite{Hernandez2020, Phillipson2021, sales2023} and many others, and therefore devising ways to efficiently find the ground states of spin glass models constitutes a goal of utmost importance for both scientific and industrial problems.

%Since nature itself is full of examples of physical systems that naturally tend to find its minimum energy state, many attempts to design physics-inspired algorithms have been proposed to this end. For instance, inspired by the process of slow thermal cooling and heating that, in the metallurgic industry, aims to obtain the most robust configuration (\textit{i.e.} the lowest energy one) of the metal one is working by controlling its free energy, 
Many physics-inspired algorithms have been devised for that end. One way is to use standard Markov chain Monte Carlo (MCMC) methods to mimic thermal fluctuations and utilise various temperatures to explore the rugged energy landscapes of spin-glass systems. Inspired by the process of slow thermal cooling and heating in the metallurgic industry, Simulated Annealing (SA) \cite{Kirkpatrick1983} was one of the first algorithms of such type to prove its value in real-life applications already several decades ago. As an evolution of SA and taking advantage of massively parallel computing platforms such as Graphics Processing Units (GPU) available nowadays, recently proposed Population Annealing (PA) \cite{Gubernatis2003, Machta2010} is an extended ensemble Monte Carlo algorithm that combines SA's thermal annealing with a population of independent replicas of the system under study. Thermal annealing is then significantly sped up by replicas population resampling procedure done according to the Gibbs distribution. For thermally equilibrated and sufficiently large replicas population, resampling allows to change population temperature in a single step while the population remains approximately equilibrated. %, thus significantly speeding up annealing. %Resampling of the population is done according to the Gibbs distribution, such that those replicas that at a given point of the simulation have found a lower energy configuration are more prompt to get replicated, while those states with higher energy tend to be eliminated. 
This %natural selection-like process 
has been found to improve, in many cases, the performance of the algorithm over SA \cite{Wang2015}. Although  PA is a sequential Monte Carlo method \cite{Wan2015B} it still relies on basic a Metropolis algorithm \cite{Metropolis1953} to independently equilibrate each of the replicas %that correspond to the population at each temperature in the annealing schedule 
after every resampling step, in order to ensure that the correlations between replicas introduced by resampling are eliminated. The Metropolis algorithm is known to slowly equilibrate, {\it e.g.} in the critical region or in case of rugged landscapes \cite{newmanb99}, in which it gets trapped by local minima. The latter is problematic in spin systems with quenched disorder at low temperatures, for example when studying three-dimensional Edwards-Anderson (EA) spin-glass systems \cite{EdwardsAnderson1975}. This fact can sometimes cause PA to encounter hard disorder instances for which population cannot be properly equilibrated \cite{Wan2015B}. One way to overcome the Metropolis' algorithm deficiencies is to use cluster moves. In case of uniform, ferromagnetic, Ising models the celebrated Swendsen-Wang \cite{PhysRevLett.58.86} and Wolff algorithms \cite{Wolff1989} work very efficiently. The case of spin-glass models is more problematic as clusters may grow fast, potentially spanning the whole lattice. The Houdayer cluster algorithm \cite{Houdayer2001} of isoenergetic moves was designed specifically for those models. It offers substantial speed up in two dimensions and it has recently been efficiently adapted in three dimensions (and other connection topologies), even though only in specific temperature and frustration windows \cite{PhysRevLett.115.077201}. Therefore the search for finding efficient cluster moves in Monte Carlo algorithms for spin-glass models, especially moves compatible with massively parallel computations, remains an important challenge.

On the other hand, while the new advances on Quantum Annealing (QA) machines \cite{FINNILA19943, Satoshi2008} pose them as firm contenders to efficiently solve these kind of QUBO problems, they're still far away from surpassing classical computation. As a counterpart, new quantum-inspired algorithms for classical computers have been recently developed, taking advantage from current dominant position of such platforms. Simulated Quantum Annealing (SQA) \cite{Crosson2016} constitutes the paradigmatic example within this field, in which a classical simulation of the adiabatic transition of a wave function from the initial state to the ground state of the spin-glass model is performed. In this case, the simulation of a coherent exploration of the Hilbert space upon changing Hamiltonian replaces the simulation of thermal annealing.

In this work we propose a new quantum-inspired method to improve the exploration of the optimization and quenched disorder spin problems in PA, by the evolution (annealing) of certain local constrains. Taking inspiration from the quantum Toric Code \cite{KITAEV2003}, the creation, movement and annihilation of topological defects within the lattice allows for the conduction of non-local moves. In two dimensions such moves can be performed on the standard spin-glass random-bond Ising model with an expanded configurational space ({\it i.e. allowing for topological defects}) and are equivalent to cluster updates. In three dimensions we propose a numerically hard model, equivalent to the random 3d random plaquette gauge model (RPGM) \cite{10.1063/1.1499754}, which expanded configurational space allows for respective topological defects and non-local moves, even though the analogy with the spins clusters is not that straightforward.
The evolution of such defect states starts with relaxed constrains (free defect creation) and is controlled externally by the increase of a field parameter added into the original Hamiltonian that progressively penalizes their proliferation. The preparation of the starting thermally equilibrated replica population is straightforward in any temperature as defect constrains are relaxed. Therefore annealing can be now conducted starting from any temperature, and can be performed in a two-dimensional parameter space (temperature and defect control parameter). In this regard we define an adaptive step procedure that is able to optimize the annealing of the system in this parameter space, gaining flexibility over PA, which is seen to increase the thermalization of the systems without compromising other related and desired properties. Importantly, the proposed method is itself a variation of PA, and is therefore massively parallelizable as well, leading to parallelizable effective non-local moves.

One should note that an efficient general solution to the numerically hard (NP-complete in the worst scenario) spin-glass and optimization problems, with increasing lattice sizes, is most probably impossible on classical computers. Yet even for systems and lattice sizes amenable for current computational resources, the hard disorder cases are the limiting factor. Different annealing methods (temperature in PA, adiabatic wave function changes in SQA, topological defect constrains in the present variation of PA) have the potential to address different hard disorder cases. Indeed that is what we observe: disorder cases that are difficult for standard PA are usually not that hard in our Defect-driven PA. 

The paper is organized as follows. In Section \ref{sec: methodology} we first give a description of PA and discuss a metric for the confidence of the solution found corresponding to the global minimum, which will eventually allow us to compare the newly proposed method. We then introduce the random 3d Ising gauge theory model in plaquette representation that we named Random Field Wall (RFW) model and describe its topological defects dynamics, to finally propose the Defect-driven Population Annealing (DPA) modification. In Section \ref{sec: results} we first discuss the hardness of the RFW model and compare it against the EA one, apply DPA to RFW lattices of linear sizes $L=4, 6, 8$, and compare the results against one of the state-the-art algorithms, PA, which, in the task of finding ground states of spin glasses, is equivalently efficient to other ones such as Parallel Tempering (PT) \cite{Wang2015}. Further, in this same work PA was found to reach the same ground states than other completely different types of algorithms, such as the Hybrid Genetic algorithm \cite{Palassini}. Finally, we conclude this work with a discussion in Section \ref{sec: conclusions and outlook}.

\section{Methodology} \label{sec: methodology}

\subsection{Population Annealing}
Population Annealing  is a generalized ensemble extension of SA in which a family of a total of $R_0$ replicas of the same system are independently simulated in parallel.  PA starts considering all $R_0$ replicas at infinite temperature, at which equilibration is easy but encountering the global ground state is difficult, and anneals the population towards a low temperature at which equilibration is difficult but the probability of finding the ground state is higher.  We denote the annealing schedule as $S_\beta=[\beta_0, ..., \beta_{N_T}]$, with $\beta$ the inverse temperature and $\beta_i<\beta_{i+1}$ $ \forall i$, where $i$ labels each  resampling and equilibrating step. Equilibration of the  replicas  is performed through a Markov Chain Monte Carlo (MCMC) method for a number of sweeps $N_{\rm{sweeps}}$, and then the population is resampled according to the Gibbs distribution, which is known to enhance the efficiency of the algorithm compared to SA. This resampling step includes elimination and proliferation of replicas that somewhat resembles the selection part of a genetic algorithm, in which replicas with a lower energy (that is, better fitness) are set to have several offspring, while those having a higher one tend to be eliminated. When resampling from an inverse temperature $\beta$ to $\beta'$, the normalized weight for replica $j$ is defined as
\begin{equation} \label{eq: resampling normalized weights PA}
    \tau_j(\beta, \beta') = R_0 \frac{e^{-(\beta'-\beta)E_j}}{\sum_r e^{-(\beta'-\beta)E_r}}, 
\end{equation}
where $R_0$ is the initial number of replicas of the system and the index $r$ runs over all replicas, and we choose its number of descendants $n_j$ as
\begin{equation} \label{eq: number of sons PA}
    n_j = \begin{cases}
    \left \lceil{\tau_j}\right \rceil  &\quad\text{with probability } P_{\rm{ceil}}=\tau_j-\left \lfloor{\tau_j}\right \rfloor \\
    \left \lfloor{\tau_j}\right \rfloor  &\quad\text{with probability } P_{\rm{floor}} = 1-P_{\rm{ceil}}.
    \end{cases}
\end{equation}
This choice of probability minimizes the variance of $n_j$ and lets the population's size vary around a mean value $R_0$ with a fluctuation of $\sqrt{R_0}$. %, where $R_i$ is the number of replicas at step $i$. 
This choice reduces the correlation between replicas in the descendant population~\cite{Wan2015B}.

An additional and important advantage of PA is that it yields an estimate of the free energy of the simulated system at no additional cost. For the annealing schedule $S_{\beta}$ described above, the estimated free energy at inverse temperature $\beta_k$, $\tilde{F}(\beta_k)$ is \cite{Ebert2022}:
\begin{equation}
    -\beta_{k}\tilde{F}(\beta_k) = N_s \ln 2 + \sum_{i=0}^{k-1}\ln Q(\beta_i, \beta_{i+1}),
\end{equation}
where $N_s$ is the total number of spins in the system and $Q(\beta, \beta')$ is the normalization factor used by PA in the computation of the resampling weights, Eq.~\eqref{eq: resampling normalized weights PA}:
\begin{equation}
    Q(\beta, \beta') = \frac{1}{R} \sum_r e^{-(\beta'-\beta)E_r}.
\end{equation}

\subsection{Confidence on the solution found}
Due to their heuristic nature, algorithms such as SA or PA may not reach the global, \textit{true} ground state of a certain model at hand. In such cases, one has to conform with a measure of the likelihood of the solution found being a global minimum, or as close as possible to it. Specifically for PA, we can assess this likelihood by means of measuring two different parameters on the final population of the algorithm's run. The first one, $g_0$, is the fraction of replicas that, in the final population, are in the same minimum energy state detected during the entire simulation, $g_0=N_0/R$, with $N_0$ counting the number of times that this state appears in the final stage. Given that PA can accurately estimate the free energy of the simulated system, $g_0$ can be estimated as well from its definition
\begin{equation} \label{eq: g0}
    g_0 = \frac{2e^{-\beta E_0}}{Z(\beta)} = 2e^{\beta(E_0-F)},
\end{equation}
with $E_0$ the ground state energy of the system. The value obtained by measuring $g_0$ from $N_0$ in the simulation should coincide with Eq.~\ref{eq: g0} in the limit of infinite replicas \cite{Wang2015}.
The second considered parameter is the effective number of surviving families $N_{\rm{eff}}$, which measures the number of replicas that have found the lowest energy state \textit{ independently}. The effective number of surviving families can be measured as
\begin{equation} \label{eq: Neff}
    N_{\rm{eff}} = \exp[S_f],
\end{equation}
with 
\begin{equation} \label{eq: Sf}
    S_f=-\sum\nu_j \log \nu_j,
\end{equation}
the family entropy and $\nu_j$ the fraction of replicas present at the end of the simulation that descend from replica $j$ in the initial population. Effectively, family entropy is a measure of the equilibration or thermalization of the sample \cite{Wan2015B}. 

In the case of PA, the only way to be sure of having found the ground state would be to use an infinitely large population or an infinitely slow annealing process. So if  the size of the final population $R$ is large enough and the same lowest energy state appears frequently  in thermal equilibrium, we are  able to confidently state that the found configuration is the global minimum. Therefore, the higher both parameters $g_0$ and $N_{\rm{eff}}$ are after a simulation, the higher confidence we might have in the solution found being the global ground state of the system. The problem in PA is then to make sure that those two parameters are high enough, and particularly the latter, since it has been seen that it sometimes encounters hard instances for which thermalization is harder, and that therefore exhibit a very low final family entropy \cite{Wan2015B, Barzegar2018}. In the following, we explore problems of different hardnesses, with a focus on those that PA finds the hardest to solve.

\subsection{Bond and wall  representations of the models and Toric code topological defects}
For simplicity, consider first a standard 2d spin lattice with only nearest neighbours interactions, defined by Hamiltonian
\begin{equation}
    H = \sum_{\langle i,j \rangle}J_{ij}s_i^z s_j^z,
\end{equation}
with periodic boundary conditions.
We can shift to its bond representation by applying the change of variable $\sigma_b^z = s_i^z s_j^z$:
\begin{equation} \label{eq: ising hamiltonian in bond representation 2d}
    H = \sum_{b}J_b\sigma_b^z.
\end{equation}

It is important to stress that, even if the Hamiltonian looks deceptively easy, it does not correspond to free particles, as it must be understood along with the constraints associated to the set of feasible configurations (see below).
In the bond picture of a spin lattice the new binary variables $\sigma_b$ represent the state of the bond between two spins, which depends on their relative alignment instead of that of the spins themselves. The bond is either said to be up (the spins connected by it have the same value, thus $\sigma_b^z=1$), or down (the spins are not aligned and $\sigma_b^z=-1$). Note that therefore a bond configuration can be translated into a spin one with up to a decision of the value of an initial spin, which corresponds to the up-down degeneracy of the spin picture.

\begin{figure}[h] 
\centering
\includegraphics[scale = .7]{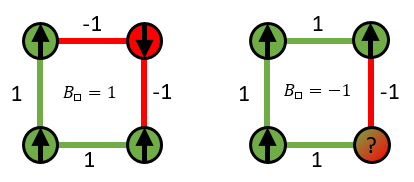}
\caption{\textbf{Physical states in the bond representation are characterized by $B_{\square} = 1$.} When a plaquette yields $B_\square=-1$, spins located on its vertices cannot be unequivocally determined and therefore the configuration does not represent a physical state of the Ising model. We thus say it contains a defect.}
\label{fig: plaquette defects}
\end{figure}

However, in contrast to the spin representation, not all $\left\{\sigma_b^z\right\}$ configurations represent a physical state. For a bond configuration to represent a physical state of the Ising model, the products of the bonds in any close path must be 1 (see Fig.~\ref{fig: plaquette defects}).  To account for this we introduce  plaquette and line operators:
\begin{equation}
    B_{\square} = \prod_{b\in\square}\sigma_b^z,
\end{equation}
\begin{equation}
    B_| = \prod_{b\in|}\sigma_b^z,
\end{equation}
where $b\in\square$ and $b\in|$ are the indices of the bonds pertaining to a given plaquette (illustrated by a square, $\square$, due to their spatial disposition) and line (illustrated by a line, $|$, due to their spatial disposition), respectively. 
Note that $B_\square$ is equivalent to TC's $B$ operator, while the spin flip is equivalent to TC's star operator, $A_s$,
\begin{equation} \label{eq: 2d spin flip}
    s_i^x=\prod_{b\in+}\sigma_b^x,
\end{equation}
where $b\in+$ are the indices of the bonds surounding spin $s_i$ (again, illustrated by a cross, $+$, due to their spatial disposition around the spin).
The condition for being able to recover a proper spin configuration translates into assuring that:
\begin{itemize}
    \item All  plaquettes yield 1.
    \item All straight lines across the lattice (both horizontal and vertical) yield 1.
\end{itemize}
If this condition is not fulfilled for a given plaquette or line, we say that it contains a (topological) defect (see Fig.~\ref{fig: plaquette defects}). In what follows we refer to such configurations as non-physical.  

\begin{figure}[h]
\centering
\includegraphics[width=\columnwidth]{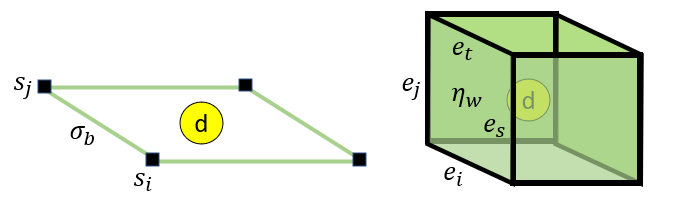}
\caption{\textbf{The 3d Random Field Wall (RFW) model is a generalization of the random 2d bond model.} By extruding a plaquette and all its elements onto an extra dimension, we obtain a cube surounded by edge spins which are 4-to-4 connected by a wall. The site enclosed inside the cube can hold a defect.}
\label{fig: plaquette to cube}
\end{figure}

Let us now shift to an appropriate 3d extension of such a 2d Ising model with topological defects in bond representation. The Kramers-Wannier dual \cite{PhysRev.60.252} to the 3d uniform Ising model is the 3d Ising gauge theory model \cite{Wegner:1971app} where spins are located on lattice edges and are subjected to plaquette interactions on each cube face. The latter model can be represented as a "wall model" where the walls are the new binary variables taking values of cube face plaquettes, the plaquette interaction $J$ is now a field acting on walls, and walls are not independent variables but are subjected to the constrain that in each cube the product of all 6 planes must equal 1. The given configuration of walls determines energy and represents all $2^N$ gauge-equivalent states (where $N$ is the number of vertices), thus in this representation entropy is greatly reduced. To visualize the 3d wall model consider an extrusion onto the new dimension of the elements of the 2d bond model described so far. This means that we add an extra dimension to all of them, and therefore degrees of freedom change as
\begin{itemize}
   \item Spins (points) become edge spins (lines).
    \item Bonds (lines) become walls (planes).
\end{itemize}
And with respect to operators (constrains)
\begin{itemize}
   \item Plaquettes (which may contain a defect) become cubes (which may contain a defect).
  %  \item Plaquettes (planes) become cubes.
    \item Lines (which may contain a defect)  become planes (which may contain a defect).
    \end{itemize}
As we previously had a total of 4 spins connected by bonds (each bond connecting two single spins) and forming a plaquette surrounded by 4 bonds that could have a defect in it, we now have 12 edge spins connected by walls (each wall connecting four edges) and all of them forming a cube surrounded by 6 walls that can contain  a defect as well (see Fig.~\ref{fig: plaquette to cube}). 
In this case, the change of variable between edge spin and wall models is
\begin{equation} \label{eq: change of variable walls to edge spins}
    \eta_w^z = e_i^z e_j^z e_s^z e_t^z,
\end{equation}
where $e_{\{i, j, s, t\}}$ are the edge spins surrounding wall $\eta_w$. 

Plaquette and line operators translate to cube and plane operators as
\begin{equation} \label{eq: cube operator}
    B_c = \prod_{w\in c}\eta_w^z,
\end{equation}
\begin{equation} \label{eq: plane operator}
    B_p = \prod_{w\in p}\eta_w^z,
\end{equation}
respectively and, again, the physicality of the lattice is imposed by $B_c=1$ $\forall c$ and $B_p=1$ $\forall p$. 
In Eqs.~\eqref{eq: cube operator} and \eqref{eq: plane operator}, $w\in c$ and $w\in p$ are the indices of the walls defining the cube $c$ and the plane $p$, respectively.

The Hamiltonian of the 3d Ising gauge model in the wall representation reads
\begin{equation} \label{eq: RFW hamiltonian without deffects}
    H = \sum_w J_w \eta_w^z.
\end{equation}
However, analogously to Eq.~\eqref{eq: ising hamiltonian in bond representation 2d}, this is not a free particle model. It has to be understood along with the constraints associated to the set of feasible configurations given by $B_c=1 \forall c$ and $B_p=1\forall p$. Of course, instead of imposing these constrains on the phase space, they can be included in the Hamiltonian through Lagrange multipliers. In following sections, we will employ a similar approach to introduce topological defects into the model.

If we now consider quenched disorder signs of plaquette interactions in 3d Ising gauge model we get the 3d random plaquette gauge model, which appeared recently in a study of stability of topological quantum memory \cite{10.1063/1.1499754} based on the TC. Following Monte Carlo simulations revealed that small concentration of negative sign plaquettes drives such system from the "Higgs phase" to the disordered "confining phase" \cite{WANG200331, OHNO2004462}. Furthermore, the phase structure of the 3d RPGM model was studied using Wu-Wang duality \cite{TAKEDA2004377}. Here we study 3d random plaquette gauge model with Gaussian disorder for plaquette interactions in the "wall" representation that we will refer to as 3d Random Field Wall (RFW) model. This model is not dual to the 3d Edwards-Anderson spin-glass model (also called random-bond Ising model), which consists of up/down spins arranged in a cubic lattice with only nearest-neighbor couplings, whose values are drawn from a normal distribution. 
In fact 3d RFW relation to the spin-glass physics is unclear. Replica symmetric mean-field solutions for random gauge theories \cite{ARAKAWA2004152} predict occurrence of "gauge-glass" phase although probably realized in higher dimensions. Still, we found the problem of finding the ground state of the 3d RFW model to be numerically hard (harder than 3d Anderson-Edwards spin-glass model for typical disorder case) and therefore suitable for testing PA and our algorithm. %\cite{Katzgraber_2013}.
Finally, we note that the original 2d spin lattice can be seen as a special case of a 3d RFW lattice with a single layer of cubes, and in which we restrict the spin and wall flips to those perpendicular to the lattice plane.

\subsection{Defects dynamics}
\begin{figure}[h]
\centering
\includegraphics[scale = .7]{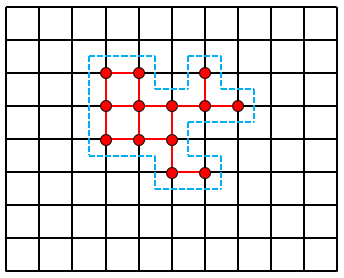}
\caption{\textbf{Allowing single-bond flips enables cluster, non-local updates in the lattice.} By applying single bond-flips, a pair of defects can be created and moved around the lattice. When they collide and annihilate, the closed path they have followed (blue dashed line) effectively realizes a non-local cluster update of the spins it encloses (red dots). The inner bonds are left unchanged in such update.}
\label{fig: cluster update}
\end{figure}

Besides the usual spin-flip movement, Eq.~\eqref{eq: 2d spin flip} (and Eq.~\eqref{eq: edge spin flip} for the 3d case, see below), we can now introduce TC topological defects into the lattice by allowing single-bond (-wall) flips with the operator $\sigma_b^x$ ($\eta_w^x$), thus extending the configurational space beyond the one representing physically feasible solutions. Note that this operator not only can create a pair of defects but can also make them move independently of each other around the lattice and eventually annihilate them if they happen to collide.
This defect propagation introduces cluster-based, non-local dynamics into the MCMC simulation: %similarly to the Wolff algorithm \cite{Wolff1989}, 
if the pair of defects created follow a longer closed path until they collide, all the spins enclosed in it are effectively updated as a cluster, since all surrounding bonds have been updated (see an example in Fig.~\ref{fig: cluster update}).

\begin{figure}[h]
\centering
\includegraphics[width=\columnwidth]{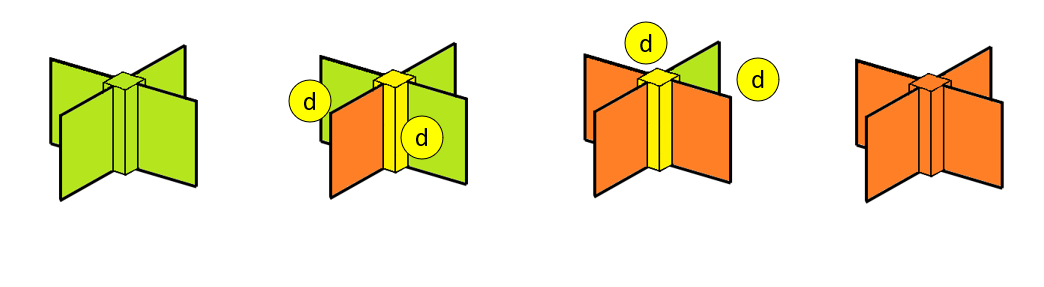}
\caption{\textbf{Graphical representation of an edge spin flip.} Green and orange colors label $\pm 1$ values for the walls and the edge spin, while yellow means that there are defects in the adjacent cubes and therefore the edge cannot be unequivocally determined. We stress that, even if only one edge spin is shown for clarity, each of the walls is surrounded by (is coupling) four edge spins and thus the shown defects must be seen as pertaining to a whole cube (for instance, they could move to the cube on its top instead).}
\label{fig: edge spin flip}
\end{figure}

From this perspective, the spin flip operator can be seen as creating a pair of defects and moving them around a given spin (vertex), see Fig~\ref{fig: edge spin flip}. Analogously to the 2d case, for the RFW the edge spin flip consists of updating all walls surrounding a given edge-spin
\begin{equation} \label{eq: edge spin flip}
    e_\mu^x = \prod_{w\in+}\eta_w^x,
\end{equation}
where $w\in+$ are indices of the walls located around the edge spin $\mu$. Analogously to the 2d case, here the notation $+$ is intended to illustrate the spatial disposition of the four bonds surrounding the spin on the vertex $\mu$.

\subsection{Defect-driven Population Annealing}
We are now ready to introduce our algorithm, which combines PA's features with the non-local cluster updates discussed before. We refer to this modification as Defect-driven Population Annealing (DPA). To this aim, we introduce an additional field parameter $\kappa$, with which we control the appearance of defects in the lattice by assigning an energy penalty to each of these generated defects. 
First of all, note that at any time we can count the number of cube and plane defects within a lattice, $N_c^{(D)}$ and $N_p^{(D)}$, with the help of operators Eq.~\eqref{eq: cube operator} and Eq.~\eqref{eq: plane operator}, as
\begin{equation}
    N_c^{(D)}=N_c-\sum_c B_c,
\end{equation}
\begin{equation}
    N_p^{(D)}=N_p-\sum_p B_p,
\end{equation}
where the summations run over all cubes and all planes in the lattice, and $N_c$ and $N_p$ are the total number of cubes and planes, respectively.
Accounting for this penalty into Eq.~\eqref{eq: RFW hamiltonian without deffects}, the final Hamiltonian of the RFW model in the  unconstrained phase space ({\it i.e.} including non-physical states) reads
\begin{equation} \label{eq: final Hamiltonian}
    H = \sum_w J_w\eta_w^z + \kappa\left[ \left( N_c-\sum_c B_c \right) + \left( N_p - \sum_p B_p \right) \right].
\end{equation}
Note that the $\kappa$ term incorporates constrains into the Hamiltonian similarly to the Langrange multiplier method. However we use it differently.
%The constraints of the RFW model Eq.~\eqref{eq: RFW hamiltonian without deffects} have been incorporated into the Hamiltonian as the $\kappa$ term. 
By initializing the simulation at $\kappa=0$ we ensure that the system can be readily thermalized at any chosen temperature and acquires a lot of defects. Then, by changing this parameter to high enough values, we make sure that the constraints are fulfilled at the final population. Therefore, at the final stages of the annealing process those configurations containing defects are penalized and rare. The configurations without defects correspond to the physical configurations of the model we want to solve.
As in PA, we run the simulation on a population of $R_0$ independent replicas, and at each step in the annealing schedule we let them evolve with an MCMC procedure to ensure thermal equilibration, after which a resampling step is also carried out. Since at every resampling step both $\beta$ and $\kappa$ are updated, the normalized weights now are
\begin{equation} \label{eq: resampling normalized weights DPA}
    \tau_j(\beta, \beta', \kappa, \kappa') = R_0 \frac{e^{-[\beta'E_j(\kappa')-\beta E_j(\kappa)]}}{\sum_r e^{-[\beta'E_r(\kappa')-\beta E_r(\kappa)]}},
\end{equation}
with $E(\kappa)=\sum_w J_w \eta_w + \kappa[N_c^{(D)}+N_p^{(D)}]$.

\subsubsection{Family entropy-preserving annealing}
\begin{figure}
\centering
 \includegraphics[width = \columnwidth]{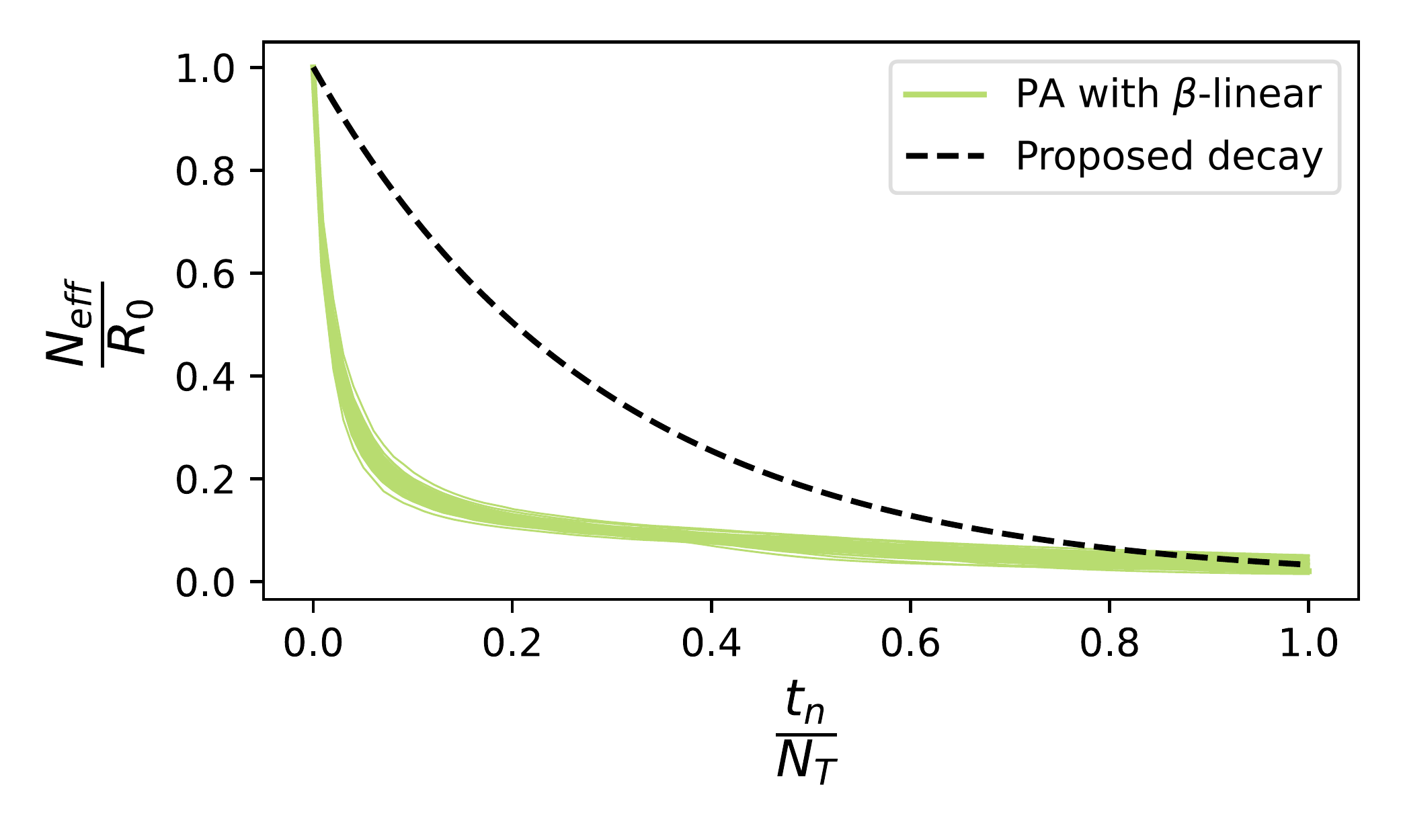}
 \caption{\textbf{Entropy-preserving annealing.} Comparison between the decay in family entropy obtained in PA simulations of $L=4$ RFW lattices with a $\beta$-linear annealing schedule (green lines) and the objective, implemented decay (black dashed line). Each green line is the decay obtained for a different disorder. Although final family entropy may seem very similar in this scale, its fluctuations in PA simulations describe differences between hard and easy disorder cases. %The implemented annealing schedule tries to loose a constant portion of families at each simulation step, while the classic one loses a huge amount at the beginning.
 }
  \label{fig: entropy preserving annealing}
\end{figure}

As have been discussed, a bad thermalization of the system is characterized by a low final value of family entropy, which is reduced everytime a resampling step is carried on the population. In standard Population Annealing, in which a linear schedule in $\beta$ is classically used in the literature \cite{Wang2015}, the family entropy tends to follow a geometric decay, Fig.~\ref{fig: entropy preserving annealing}. However, the difference between hard and easy disorder cases manifests itself in subtle changes in the decay line and, importantly, its end value. Since our Defect-driven Population Annealing algorithm utilizes annealing in a two-parameter space, the temperature and the $\kappa$, we use a different schedule. %Instead,
In order to maximize familly entropy value during the whole simulation, we propose to adapt the annealing process such that, at each resampling step, a constant portion of the number of surviving families is lost. This way the family entropy follows an exponential decay towards an objective value, which we can tune externally. To this end we implement an adaptive step procedure for both annealing schedules, $S_{\kappa}$ and $S_{\beta}$. Before each resampling step $(\beta, \kappa) \to (\beta', \kappa')$, we compute the optimal values of $\kappa'$ and $\beta'$ needed in order for the desired portion of families to be lost lost, such that the obtained a family entropy follows the specified decay
\begin{equation}
    \frac{N_{\rm{eff}}(t_n)}{R_0} = \left(\frac{N_{\rm{eff}}^{\rm{desired}}}{R_0} \right)^{\frac{t_n}{N_T}},
\end{equation}
where $t_n$ is the simulation step, $N_{\rm{eff}}(t_n)$ is the number of surviving families at that time, and $N_{\rm{eff}}^{\rm{desired}}$ is the objective number of surviving families at the end of the simulation.

%One can intuitively see that $g_0$ and the final $N_{\rm{eff}}$ can be somewhat interchanged, as each resampling step may eliminate replicas with high energy values from independent families in order to replicate a more optimal one. As we shall see later, the drive through the two-dimensional $(\beta, \kappa)$-space allows the system to minimize the loss of surviving families, yielding values on the upper section of the range of the blue lines in Fig.~\ref{fig: entropy preserving annealing}, while still maximizing the final $g_0$. This will happen to be achievable even for these hardest cases for which PA does not properly equilibrate the population and thus yields a negligible family entropy. However, if a too ambitious final objective family entropy is set, it might come at the cost of a large loss of $g_0$. In that case, an extra resampling step at a very low temperature can be applied, such that we trade one for the other by effectively replicating those states with the lowest energy.

\subsubsection{Remaining defects}
It may happen that only annealing through $\kappa$ is not enough to ensure that we get rid of the defects at the end of the simulation and that some of them get stuck in potential traps, thus ending up with physically unfeasible lattices. In such cases two additional procedures can be applied. On the one hand, an N-fold way algorithm \cite{BORTZ1975} among the already existing defects only (thus they can be moved and annihilated, but no new defects can be created) is used to increase their mobility and speed up their dynamics. On the other hand, an attraction potential between defects can be used to reduce the probability of them getting stuck inside potential traps while forcing them to collide. This is done by the introduction  into Hamiltonian Eq.~\eqref{eq: final Hamiltonian} of the additional term
\begin{equation} \label{eq: gravitational attraction}
     H_G = G \sum_{i=1}^{N_{\rm{def}}}{\sum_{j=0}^{i-1}{\frac{1}{D_{ij}^{\alpha}} } },
\end{equation}
where $G$ is a negative constant that can be tuned, $N_{\rm{def}}$ is the number of defects present in the lattice, $D_{ij}$ is the euclidean distance  between defects at sites $i$ and $j$, and $\alpha$ can be used to make the interaction potential shorter or longer ranged.

\subsubsection{Outline of the algorithm}
To conclude, in this section we present a pseudocode of the Defect-driven Population Annealing algorithm.

\begin{algorithm}
\caption{Defect-driven Population Annealing}
    \begin{algorithmic}[1]
    \For{each replica r}
        \State r $\gets$ initialize
    \EndFor
    \For{$\rm{i_{\beta}}$ in $\rm{N_{\beta}}$}
        \State{$\rm{N_{eff}^{(desired)}} \gets ProposedDecay[i_\beta]$}
        \State{Estimate $(\beta', \kappa')$ such that $\rm{N_{eff}[i_\beta] \approx N_{eff}^{(desired)}}$}
        \State Resample$(\beta, \beta', \kappa, \kappa')$
        \For{each replica r}
            \For{sweep in $\rm{N_{sweeps}}$}
                \For{n in $\rm{N_{spins}}$}
                    \State $\rm{s} \gets U(0, \rm{N_{spins}})$
                    \State Metropolis$(\rm{r}, \rm{s}, \beta', \kappa')$
                    \If{$\rm{i_\beta} < i_\beta^{(critic)}$}
                        \State $\rm{w} \gets U(0, \rm{N_{walls}})$
                        \State Metropolis(r, w, $\beta'$, $\kappa'$) 
                    \Else
                        \If{$\rm{N_{defects}(r) \neq 0}$}
                            \State{NFoldWay$(\rm{r}, \beta', \kappa')$}
                        \Else
                            \State $\rm{w} \gets U(0, \rm{N_{walls}})$
                            \State Metropolis(r, w, $\beta'$, $\kappa'$)
                        \EndIf   
                    \EndIf
                \EndFor
            \EndFor
        \EndFor
        \State $\beta \gets \beta'$
        \State $\kappa \gets \kappa'$
    \EndFor
    \end{algorithmic}
\end{algorithm}

As in regular PA, the algorithm starts by randomly initializing the configurations of each replica in the population. 
The annealing schedules, $S_\beta$ and $S_\kappa$, last a total of $N_\beta$ steps, and then for each annealing step the algorithm estimates the next point in the schedules, $(\beta', \kappa')$ such that the family entropy obtained from the resampling follows the proposed exponential decay. 
After resampling, each replica $r$ is thermalized with a total number of sweeps $N_{\rm{sweeps}}$ at $(\beta', \kappa')$. 
To this end, a number of $N_{\rm{spins}}$ spin updates and wall updates are proposed and accepted with a Metropolis algorithm. 
In case the simulation is reaching its end ($i_\beta \geq i_\beta^{\rm{(critic)}}$), and a given lattice still has sites containing topological defects, the dynamics are accelerated with an NFoldWay algorithm instead of using a regular Metropolis one, such that no more defects can be created but only moved around the lattice (with the additional gravitational term, Eq.\eqref{eq: gravitational attraction}) so that their probability of closing a path and annihilating increases.
After all replicas are thermalized, the next annealing steps starts until the schedules reach its maximum length $N_\beta$.

\section{Results} \label{sec: results}

\subsection{Hardness of the RFW model}
\begin{figure}
\centering
 \includegraphics[width = \columnwidth]{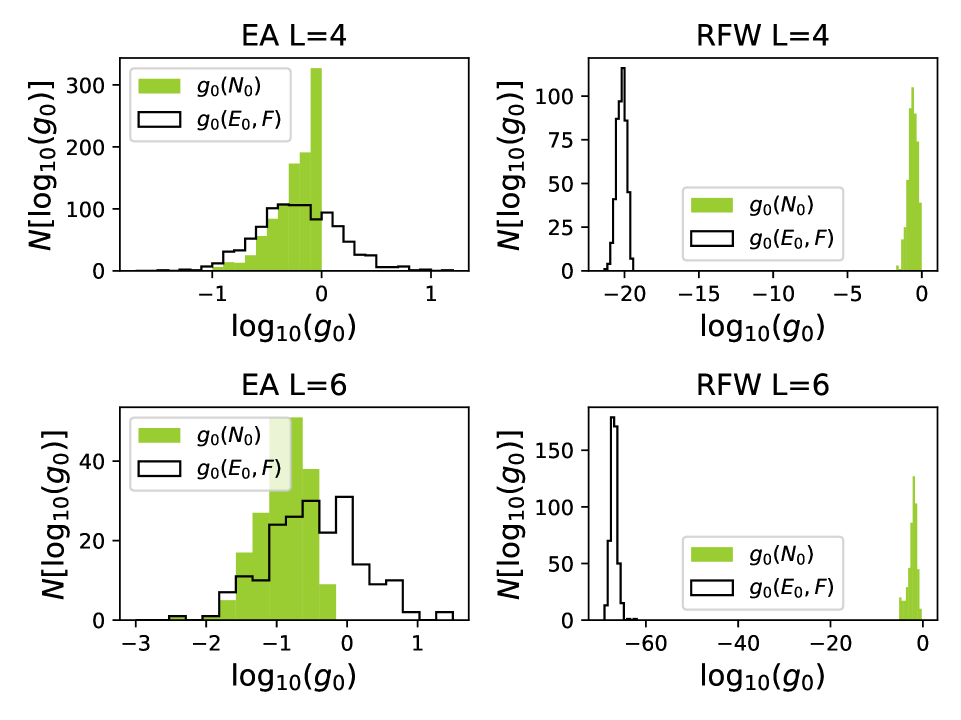}
 \caption{\textbf{RFW models are computationally harder than their EA counterparts.} Histograms of the values of $g_0$ obtained with the two discussed independent measures, for various disorder instances of EA (left panels) and WRF (right panels), of sizes $L=4$ (top panels) and $L=6$ (bottom panels).}
  \label{fig: g0 EA v RFW}
\end{figure}

One of the standard and most widely used spin glass models is the so-called Edwards-Anderson model \cite{EdwardsAnderson1975}, for which reliable results on its 3-dimensional version have been reported for sizes of up to $L=10$ \cite{Wang2015}, or equivalently, a total of $N_s=1000$ spins. Bigger lattice sizes have also been explored in the literature, but with a lower degree of confidence on the quality of the thermalization achieved \cite{Wan2015B}. On the other hand, reducing the frustration of the problem solved by considering easier disorder distributions can indeed allow for the exploration of larger lattices. For example, by considering 3d lattices with nearest neighbours and couplings distributed uniformly as $J=\pm1$, which in effect greatly reduces the spin frustration and has the potential of exponentially increasing the degeneracy of the ground state, results for sizes of up to $L=40$ have been obtained \cite{Baity_L40}. %It consists of up/down spins arranged in a cubic lattice with only nearest-neighbor couplings, whose values are drawn from a normal distribution. 
As discussed earlier, the RFW model %is an non-standard 3d extension of the 2d-EA lattice that 
does not correspond to the standard 3d-EA model but allows for the effective dynamics of topological defects. It is well known that multi-spin interacting models (with \textit{multi} referring to more than two) can in general be very difficult to sample and optimize.
The RFW model, Eq.~\eqref{eq: RFW hamiltonian without deffects}, is equivalent to 3d random Ising gauge model with four-spin plaquette interactions, so indeed can be expected to be difficult. On the other hand, the form of the interactions depends on the representation. The 2d Edwards-Anderson model, Eq.~\eqref{eq: ising hamiltonian in bond representation 2d}, which lacks hard instances, looks like a free-particle model with four-particle constrains in the bond representation. Similarly the RFW model in the wall representation looks like a free-particle model with six-particle constrains.
Furthermore the 2d EA model in bond representation, can be seen as a special case of a 3d RFW lattice with a single layer of cubes, and in which we restrict the spin and wall flips to those perpendicular to the lattice plane. 
%In that representations both models have multi-particle constrains, which can be imposed on the phase space or added as Lagrange multipliers to the Hamiltonian.
Because of that, the multi-spin nature of the constrains in the wall representation of the RFW model %Eq.~\eqref{eq: RFW hamiltonian without deffects} 
is not obvious argument that RFW model is more difficult to simulate than already hard 3d EA model. Also there are not many literature results about it.
Therefore in this section we explore hardness of the RFW model, given by Eq.~\eqref{eq: RFW hamiltonian without deffects}, by comparing it against the standard 3d EA one. The RFW model is, as we show below, much more difficult to solve for typical disorder cases %as its energy landscape appears to be considerably more difficult to explore, and therefore 
as much more computational resources must be used in order to properly thermalize it. 
This difference in computational hardness can be seen straightforwardly when standard PA is used in the two models. Since PA yields independent measures for both $g_0$ and $F$, if the system is truly equilibrated their values should be properly related through Eq.~\ref{eq: g0}. We solve several instances of EA and RFW lattices of sizes $L=4$ and $L=6$ and measure $g_0(N_0)$ from the final population and $g_0(E_0, F)$ from the estimated free energy, using the same amount of computational resources for both models at each size. We observe how, while the measures of $g_0$ for the EA lattices are mostly similar in both studied sizes, RFW lattices yield much poorer results, Fig. \ref{fig: g0 EA v RFW}. Further, the difference between $g_0(N_0)$ and $g_0(E_0, F)$ appears to increase for larger RFW lattices but not for EA ones, hinting that the increase in hardness with size is more dramatic in the former model. In what follows, if the contrary is not specified the presented results for $g_0$ are obtained by measuring $g_0(N_0)$.

\begin{figure}
\centering
 \includegraphics[width = \columnwidth]{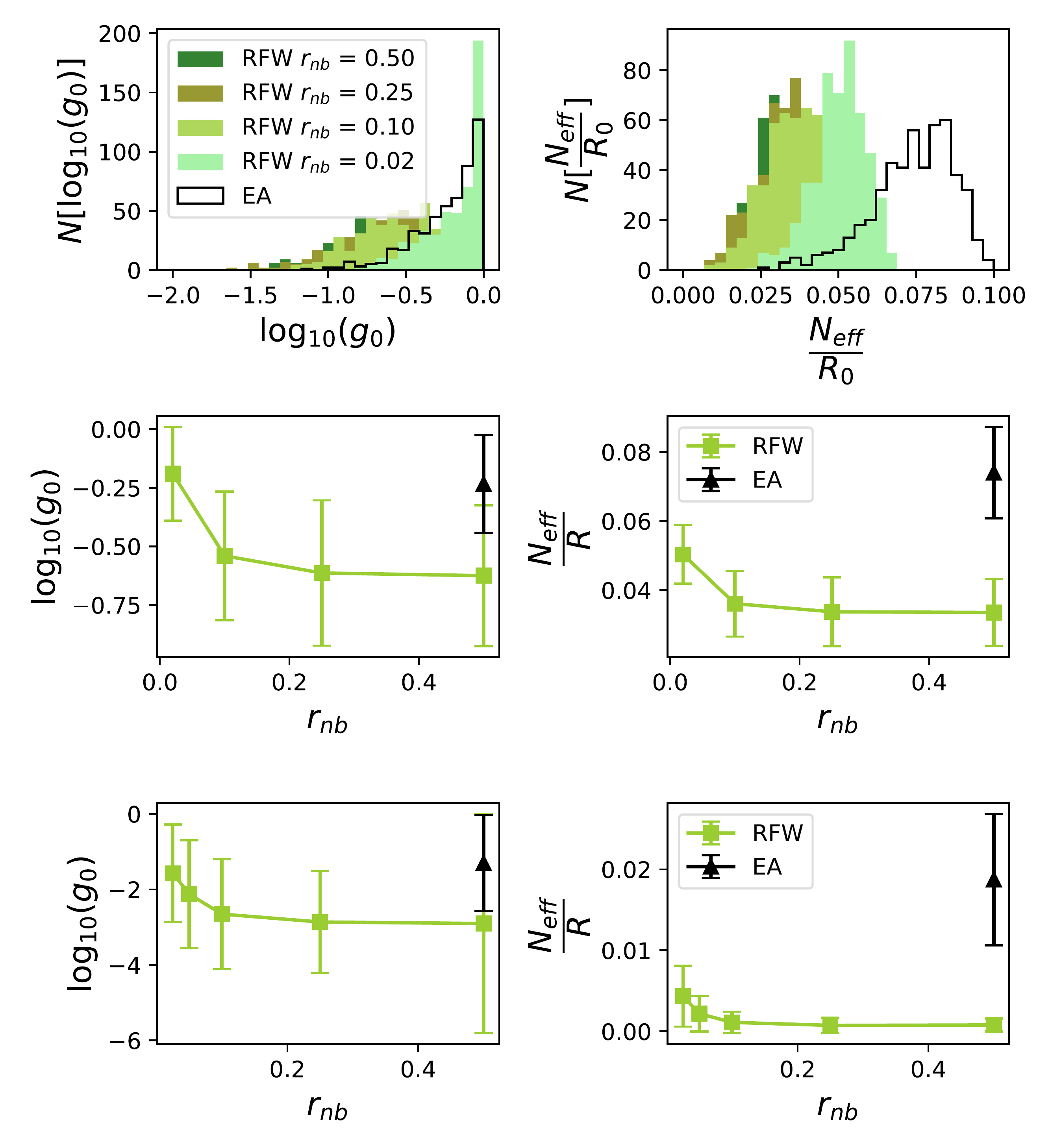}
 \caption{\textbf{The hardness of RFW models against EA clearly emerges when varying the frustration level of the solved lattices.} Top panels: histograms of $g_0$ and $\rfrac{N_{\rm{eff}}}{R_0}$ obtained for $L=4$ RFW lattices of various ratio of negative bonds, $r_{nb}$, along with the ones obtained for Edwards-Anderson lattices. Center panels: mean values and standard deviations of such histograms, as a function of $r_{nb}$. Bottom panels: mean values with standard deviations of the same histograms but for $L=6$ lattices. All simulations are obtained with PA with the same set of parameters. Legends in left and right panels are the same but are not shown to improve visibility of the plots.}
  \label{fig: different negative ratios}
\end{figure}

Another way we can take a look at the comparably much greater difficulty of the RFW model is by reducing the spin frustration created by the disorder by tuning the ratio of negative-valued bonds it contains. To do so, we generate normal Gaussian distributions, take their absolute value and change the sign of a randomly selected portion of them. The lower the ratio of negative bonds that the disorder instance has, the lesser the amount of frustrated spins it can potentially have, and thus the easier it will be to find its minimum energy configuration, eventually reaching the trivial limit of a ferromagnetic disorder. We consider three different ratios of negative bonds, $r_{nb}=\{ 0.25, 0.10, 0.02 \}$ and solve a total of 1000 disorder instances of RFW lattices of size $L=4$ for each of them with PA. In the top panels of Fig.~\ref{fig: different negative ratios} we show the histograms of the measured parameters discussed in previous sections, $g_0$ and $\rfrac{N_{\rm{eff}}}{R_0}$ for such cases, along with those obtained for unbiased Gaussian disorders (equivalent to $r_{nb} = 0.5$) and those obtained for $L=4$ lattices under the Edwards-Anderson model. Again, we use the same parameters for all of them. In the center panels of Fig.~\ref{fig: different negative ratios} we plot the dependence of the mean values of such histograms with $r_{nb}$, with error bars showing its standard deviations. We observe how the histograms tend to shift to higher values for smaller $r_{nb}$ and therefore for easier disorders. 

We highlight two main consequences from the results shown in Fig.~\ref{fig: different negative ratios}. On the one hand, we confirm that, as discussed earlier, higher values of $g_0$ and the family entropy are correlated with higher confidence in the solution found being the global ground state and, therefore, can be used as a metric to this end. On the other hand, it shows how strikingly harder the RFW model is compared to typical Edwards-Anderson instances with uncorrelated disorders, obtaining comparable values of $g_0$ only when restricting the frustration of disorders to a ratio of negative bonds of about $2\%$, while still getting much worse values for the family entropy. Looking at the same curves for $L=6$, we observe pretty similar results, bottom panels of Fig.~\ref{fig: different negative ratios}. Although the $g_0$ obtained for the EA model is approximately equal to the one obtained for RFW with about $2\%$ negative bonds, the family entropy is still far from it even for such low-frustration cases. It is also worth noting that, for $L=6$, this difference is even greater than for $L=4$.

Further, we implement $L=8$ RFW lattices with Gaussian disorder ($r_{nb}=0.5$), and attempt to solve them with standard PA using a large amount of computational resources. For equally large EA lattices, the parameter set $(R_0, N_T, N_{\rm{sweeps}})=(5 \cdot 10^5, 200, 10)$ is reported to be \textit{more-than-adequate resources} in \cite{Wang2015}, where $R_0$ is the initial population size, $N_T$ is the number of temperatures in the annealing schedule and $N_{\rm{sweeps}}$ is the number of sweeps carried on per temperature (each sweep consists of attempting $N_{\rm{spins}}$ MC updates on the lattice). For comparison, in our simulations we used $(R_0, N_T, N_{\rm{sweeps}})=(2.6 \cdot 10^5, 500, 20)$ and repeatedly solved 20 different disorder realizations in order to see how many times the algorithm converged to the same minimum energy state. 
Among all of them, the convergence ratios between independent runs varied from a minimum of $0\%$ (in 11 out of the 20 studied disorders) to a maximum of $12\%$ (in only three of them).

In light of these conclusions about the hardness of the present model, in what follows we restrict ourselves to a thorough study of RFW lattices of $L=4$ and $L=6$ with purely Gaussian disorders (this is, $r_{nb} = 0.5$).
Finally, we present some preliminary results for $L=8$ lattices as well. As the exposed results suggest, the strict equilibration of RFW lattices constitutes a problem out of our current computational capabilities for most of the cases. We therefore stress that we do not focus on achieving a proper equilibration of the lattices \textit{per se}, but rather on its use as a measure of the confidence of the solution found being the global ground state of the system.

\subsection{Comparison between PA and DPA}
In order to gain an insight into how the different algorithms perform, we solve, for each lattice size, several disorders with couplings randomly drawn from a Gaussian distribution.
When we talk about PA, we explicitly work with Eq.~\eqref{eq: RFW hamiltonian without deffects}, which has the constraints built in within the phase space. On the other hand, when we talk about DPA, we work with Eq.~\eqref{eq: final Hamiltonian}, which allows for the relaxation of the constraints and thus the introduction of topological defects through the $\kappa$ term.
For each algorithm we will compare the obtained $g_0$ and $N_{\rm{eff}}$.
However, we do not use the comparison between $g_0(N_0)$ and $g_0(E_0, F)$ as a metric for the equilibration of the system simulated with the two algorithms. This is becasue DPA effectively explores a larger phase space at the initial stages of the simulation, and the amount of replicas needed to accurately estimate free energy is larger. This is further exacerbated as equilibration is generally poor for RFW model as seen in Fig. \ref{fig: g0 EA v RFW}. This implies that for a given population size DPA can yield worse $F$ estimations than PA. Instead, we directly compare the obtained $g_0$ and $N_{\rm{eff}}$ as a metrics for a confidence in the ground state solution.
For a fair comparison, we should spend the same amount of computational work $W$ on both PA and DPA simulations. For that, we define the computational work $W$ as the total number of MCMC updates attempted during a simulation, namely 
\begin{equation}\label{eq: computational power}
    W = N_{\rm{sweeps}} N_T R_0
\end{equation}
with $N_T$ the total number of steps in the annealing schedule, $N_{\rm{sweeps}}$ the number of sweeps per temperature, and $R_0$ the initial number of replicas. Note that, as discussed above and according to Eq.~\eqref{eq: number of sons PA}, the population size at step $i$, $R_i$, fluctuates during the simulation and therefore a more accurate formula would be $W = N_{\rm{therm}}\sum_{i=0}^{N_T}R_i$. However, since we use a resampling protocol that minimizes fluctuations in population size, we can accurately approximate $R_0\approx N_T^{-1}\sum R_i$ and use the former. It should also be noted that the contribution of the resampling steps to the total computational work can be neglected. To study a fair comparison between both algorithms, we always use the same initial number of replicas and the same number of temperatures in the annealing schedules. In PA we implement the widely used linear in inverse temperature schedule, $S_{\beta}=[0, 5]$. 
Recently more optimal annealing schemes such as using a culling fraction have been proposed, \cite{Amey2018}, but we restrict the present study to the classical one for simplicity of implementation, as the main feature of DPA resides on the introduction of the non-local moves rather than the annealing scheme itself. To further compare both algorithms, we also implement the entropy-preserving steps in PA in the last section.
For the DPA simulations, after an exploration of the parameter space we found $(\beta_0, \kappa_0)=(2, 0)$ to be, generally, a good starting point for the entropy-preserving adaptive steps in the $L=4$ and $L=6$ cases. For $L=8$ we used $(\beta_0, \kappa_0) = (2, 0)$ as well, even though due to long simulation times the exploration of the parameter space was not that extensive and thus this initial point is more prompt to be optimized. Our method is therefore able to thermalize the system at finite temperatures as long as $\kappa_0 = 0$, which shows it has a better plasticity over PA.

Nevertheless, since in DPA we perform one spin flip and one wall flip per iteration, we distinguish between two different measures of computational work. In the first scenario, assuming that random number generation is the main bottleneck in MC algorithms, a strictly fair comparison would impose that $N_{\rm{sweeps}}^{(PA)} = 2 N_{\rm{sweeps}}^{(DPA)}$, even if this evidently yields a much poorer thermalization for the former. On the other hand, if a fast enough random number generator is assumed, one can directly count the computational work as the total number of MCMC updates attempted along the simulation. As we have discussed previously, each spin flip consists of four wall flips and, conversely, each wall flip can be counted as one fourth of a spin flip. This means that, in this case, to use the same $W$ one has to impose $N_{\rm{sweeps}}^{P(A)} = \frac{5}{4}N_{\rm{sweeps}}^{(DPA)}$ (recall that each thermalization step in DPA consists of one spin flip and one wall flip, which equate to five walls being updated). We address and compare both scenarios using $N_{\rm{sweeps}}^{(PA)}=10$ and $N_{\rm{sweeps}}^{(DPA)} = 5$ for the first and $N_{\rm{sweeps}}^{(PA)}=10$ and $N_{\rm{sweeps}}^{(DPA)} = 8$ for the second.

\subsubsection{$L=4$}
\begin{figure}[h]
\centering
 \includegraphics[width = \columnwidth]{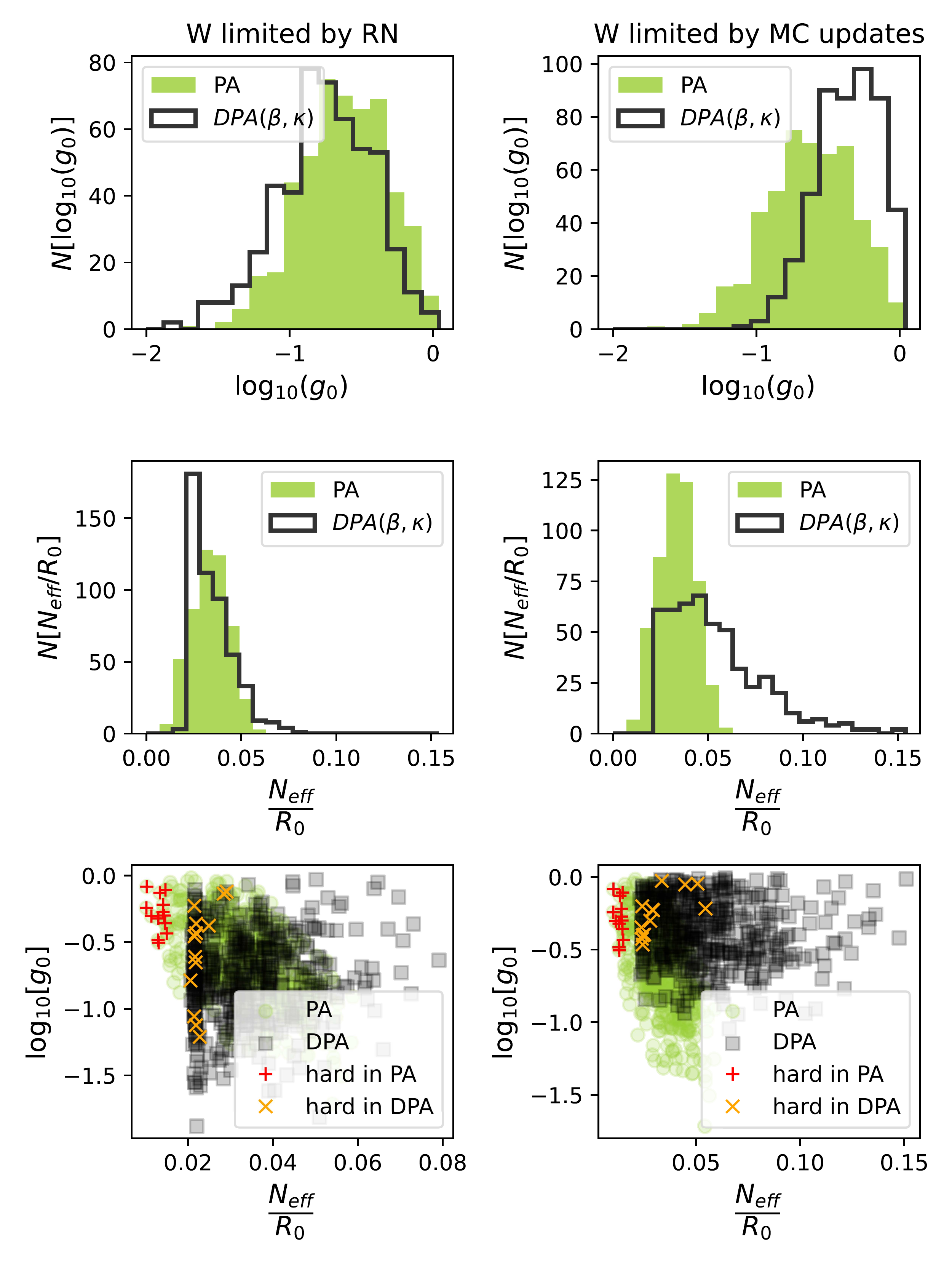}
 \caption{\textbf{DPA outperforms PA for L=4 lattices when MCMC process simulations are the computational bottleneck.} Histograms of $g_0$ and $N_{\rm{eff}}/R$ obtained by solving, by both PA and DPA, a total of $N_{\rm{dis}} = 500$ different disorders of $L=4$ RFW lattices, with bonds randomly distributed according to a normal distribution. Simulations using the same amount of random numbers (left panels) and of MCMC updates (right panels). Crosses in the bottom panels mark these instances PA finds the hardest (red), and how these same instances score in DPA (orange).}
  \label{fig: L4a}
\end{figure}

We first apply the new method to the study of $L=4$ RFW lattices and take a look at the histograms of the parameters $g_0$ and $N_{\rm{eff}}$ obtained for 500 different disorders, comparing the solutions obtained for the same disorders with a regular PA algorithm. We plot our results in Fig.~\ref{fig: L4a}, the left panels corresponding to the first scenario discussed ($W$ limited by the amount of random numbers) and the right panels corresponding to the second ($W$ limited by the number of MC updates). The top graphs show the histograms for $g_0$, the middle graphs the histograms for $\rfrac{N_{\rm{eff}}}{R_0}$, and the bottom graphs the scatter plot relating both parameters for each of the solved disorders. In these bottom panels, we also mark the disorders that regular PA finds to be the hardest to solve, identified as those obtaining the lowest values of family entropy (red crosses) and how these same disorders score when solved by DPA (black crosses). In this case, we consider a disorder to be hard whenever PA is not able to reach a final effective number of surviving families $\rfrac{N_{\rm{eff}}}{R_0} = 0.015$ or, equivalently, about 2$\%$ of all instances of disorder.

%\begin{figure}[h]
%\centering
% \includegraphics[width = \columnwidth]{images/L4a_sameRN.eps}
% \includegraphics[width = \columnwidth]{images/L4a_sameMC.eps}
% \caption{Histograms of $g_0$ and $N_{eff}/R$ obtained by solving, by both PA and DPA, a total of $N_{dis} = 1000$ different disorders of $L=4$ RFW lattices, with bonds randomly distributed according to a normal distribution. Simulations using the same amount of random numbers (top) and of MCMC updates (bottom).}
%  \label{fig: L4a}
%\end{figure}

In the first scenario (Fig.~\ref{fig: L4a}, left panels) we observe a certain trade-off between parameters (Fig.~\ref{fig: L4a}, top left and center left panels), resulting in an equivalent performance of the two methods. The fact that the adaptive steps procedure over the two control parameter space ($\beta$, $\kappa$) is able to properly drive the evolution of the system and set an objective cutoff family entropy is nevertheless noticeable (Fig.~\ref{fig: L4a}, center left panel). Also, the fact that the the family entropy is no worse than for PA is relevant, taking into account that, in this scenario, the total amount of spin updates is lower and thus one might expect a worse thermalization. With the cut-off imposed on family entropy, scores on this parameter are increased even for hard cases, but they seem to lay, generally, in the lower range for DPA as well (Fig.~\ref{fig: L4a}, bottom left panel). 

On the other hand, in the second scenario (Fig.~\ref{fig: L4a}, right panels) we see a substantial improvement with the DPA method, as the histograms for both measured parameters seem to be shifted towards bigger values and thus imply better metrics than those obtained with standard PA (Fig.~\ref{fig: L4a}, top right and center right panels). As can be seen in the bottom right panel, even some of the hard cases' metrics are improved, as they manage to escape from the lower range of family entropy without diminishing its $g_0$ score.

\begin{figure}[h]
 \includegraphics[width= \columnwidth]{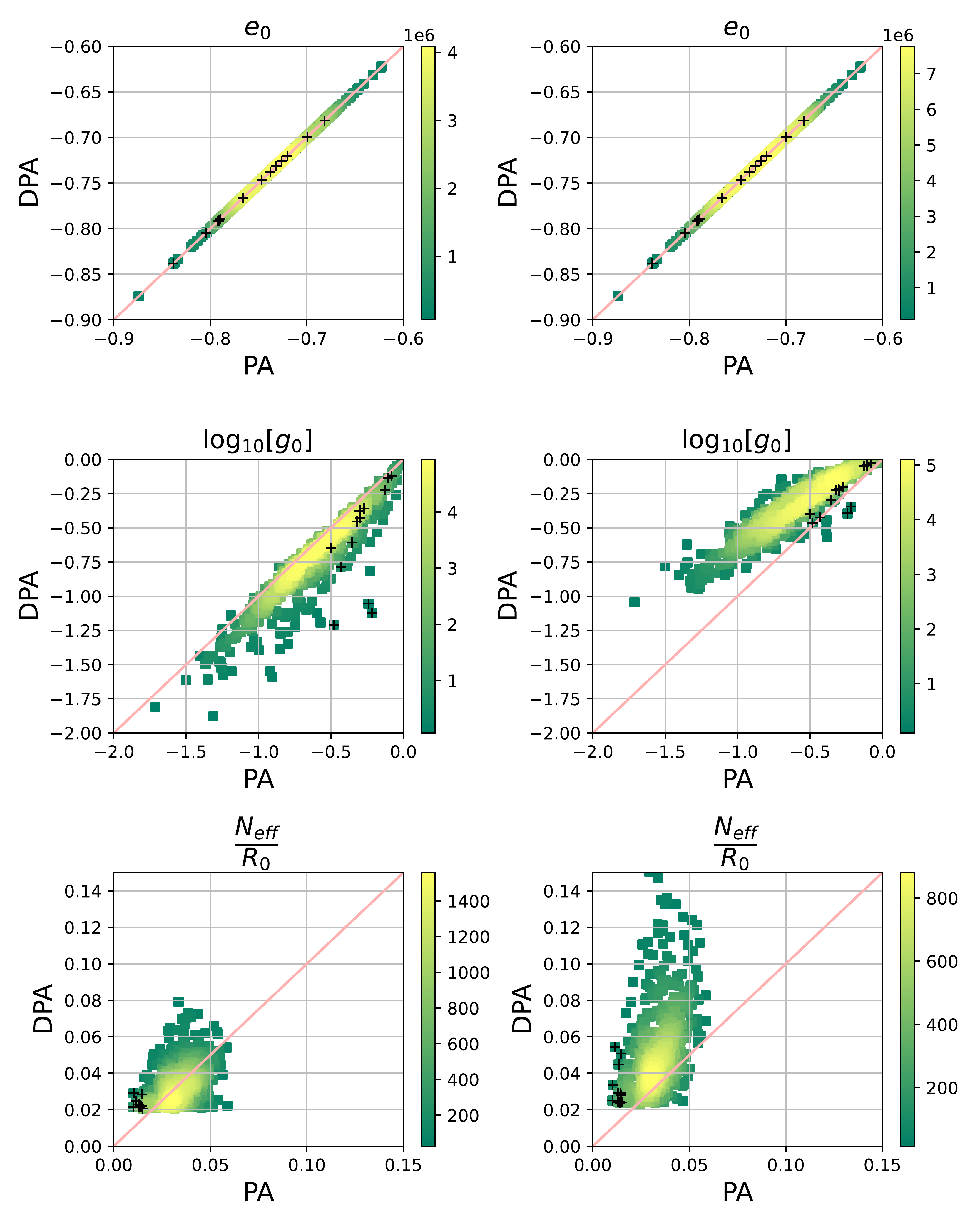}
 \caption{\textbf{DPA provides on average higher family entropies than PA when solving $L=4$ lattices.} Comparison of results obtained by PA and DPA for $L=4$ RFW lattices for the three measured parameters. From top to bottom: ground state energy, $g_0$ and effective number of surviving families. Simulations using the same amount of random numbers (left) and of MCMC updates (right). Black crosses mark these instances that PA finds the hardest.}
  \label{fig: L4b}
\end{figure}

In Fig.~\ref{fig: L4b} we take a closer look at the results obtained by the two methods, by comparing the measured parameters obtained by both for each of the solved disorder realizations. A brighter color indicates a higher density of points and the red straight line is plotted to indicate the region in which both methods yield the same results for a given parameter. We also marked those disorder instances classically labeled as hard for PA with red crosses. Again, the left and right panels correspond to the first and second scenarios, respectively.

%\begin{figure*}[ht]
% \includegraphics[width=2 \columnwidth]{images/L4b_sameRN.eps}
% \includegraphics[width=2 \columnwidth]{images/L4b_sameMC.eps}
% \caption{Comparison of results obtained by PA and DPA for $L=4$ RFW lattices for the three measured parameters. From left to right: ground state energy, $g_0$ and effective number of surviving families. Simulations using the same amount of numbers (top) and of MCMC updates (bottom).}
%  \label{fig: L4b}
%\end{figure*}

Most importantly, as can be seen in the top panels of Fig.~\ref{fig: L4b}, both algorithms find the same ground state energy for all disorders. The distribution of $g_0$ and $N_{\rm{eff}}$ is shown in the center and bottom panels, respectively. In them we can see that for the same amount of RN consumed (first scenario, left panels) the obtained $g_0$ is slightly higher for PA but certainly comparable to DPA for the vast majority of disorders (Fig.~\ref{fig: L4b}, middle left panel), even for most of the hardest instances. On the other hand, the family entropy is better with the new method (Fig.~\ref{fig: L4b}, bottom left panel), which implies that a better thermalization of the replicas is achieved (again, recall that the number of total spin updates is lower, and thus this could have been expected to be lower as well). In the second scenario, DPA obtains clearly better results than PA for both parameters, $g_0$ (Fig.~\ref{fig: L4b}, middle right panel), and $N_{\rm{eff}}$ (Fig.~\ref{fig: L4b}, bottom right panel). 

\begin{figure}[ht]
\centering
 \includegraphics[scale = .23]{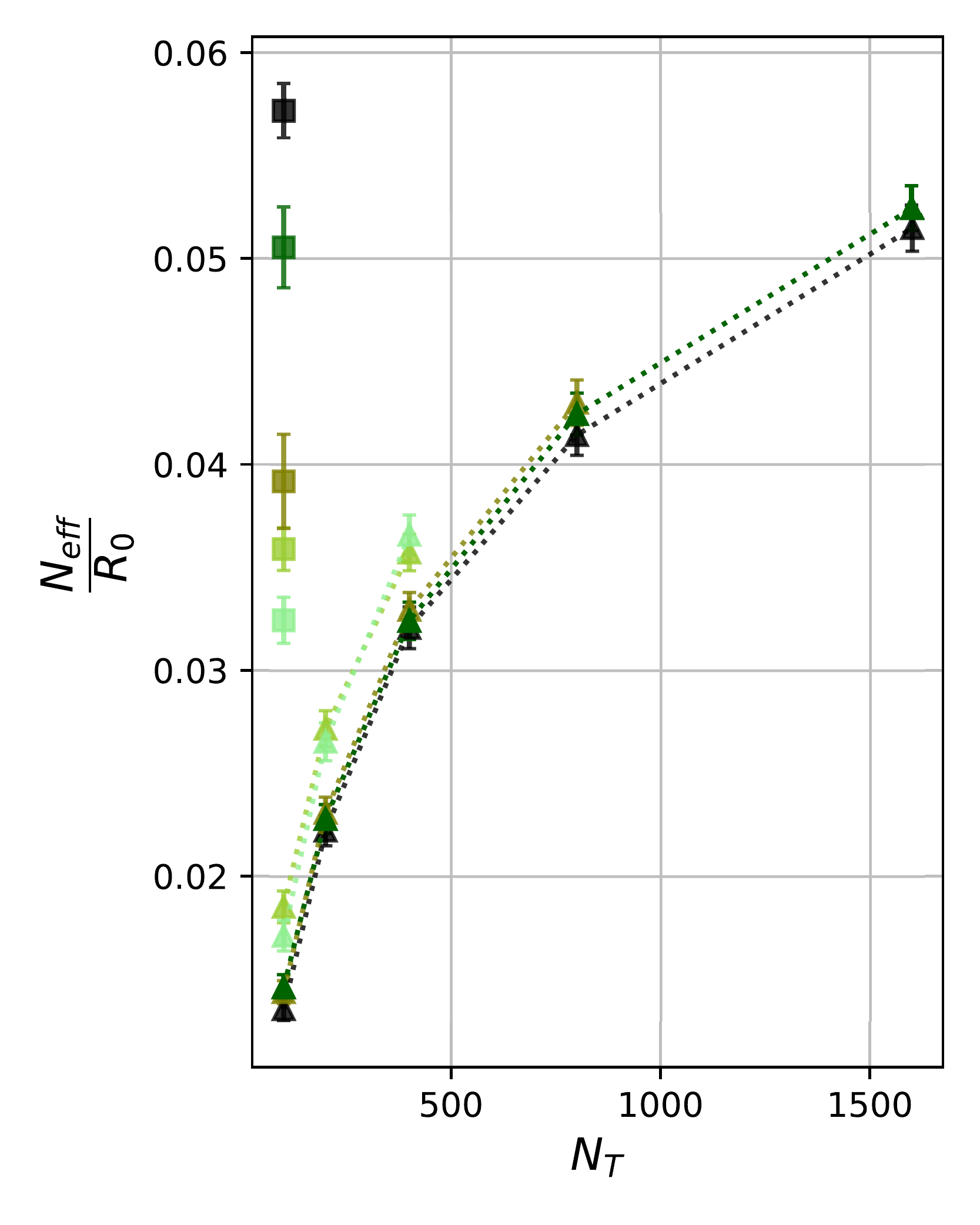}
 \caption{\textbf{Hard cases on PA require more adiabatic annealing processes to reach an equivaent family entropy to DPA.} Study of the amount of computational power that would have to be invested to solve $L=4$ RFW lattices with PA, in order to obtain a similar final family entropy as with DPA, for hard instance for which $g_0^{PA} \approx g_0^{DPA}$. Error bars are standard deviations over $N_{\rm{rep}} = 100$ repetitions of the same process at each value of $N_T$. Square markers are obtained with DPA and triangle ones with PA using an increasing number of temperature steps. Each color indicates a different disorder.}
  \label{fig: different NT, L=4}
\end{figure}

Lastly, we focus on cases of disorder realizations that are hard for the PA algorithm, that is these cases resulting in low family entropy. From Eq.~\eqref{eq: resampling normalized weights PA} follows that a possible way to overcome the difficulty of equilibration and improving the final family entropy obtained by PA is to reduce the differences $\Delta\beta=\beta' - \beta$ by increasing the number of temperature steps. %, so that the expected number of descendants for each replica approaches $1$ in a more independent way of its corresponding energy and therefore the number of eliminated families is reduced. This can be achieved by investing more computational power, Eq.\ref{eq: computational power}, by the $N_T$ side, \textit{i.e.} by using a larger number of steps in the annealing schedules within the same temperature range. This is, of course, in agreement with the intuition that, if we take smaller and smaller temperature steps, then it is easier to equilibrate the replicas given that the previous population was already in equilibrium. Taking into account that the initial population was simulated at an infinite temperature, $\beta=0$, and was therefore guaranteed to be in equilibrium \JRMS{Why?}, by induction it will therefore be the final one in the limit of $\Delta\beta \rightarrow 0$. Furthermore, in accordance with the adiabatic theorem [ref], the whole annealing process is slower and therefore the probability of finding the ground state is higher (eventually $1$ for small enough $\Delta\beta$), which is exhibited, as discussed earlier, by a higher family entropy. 
Following this idea, to further contrast both methods, we now take a collection of disorders that PA finds difficult to equilibrate and for which a very similar $g_0$ is obtained for PA and DPA, so that they lay close to the $g_0^{PA} = g_0^{DPA}$ red line in the center panels of Fig.~\ref{fig: L4b}, and repeatedly solve them several times with an increasing number of temperature steps in the PA annealing schedule (\textit{i.e.} more adiabatically). For the cases at hand, we find that for PA to obtain a family entropy similar to that obtained by DPA, we would have to invest approximately between 4 and 16 times as much computational power depending on the disorder instance, Fig.~\ref{fig: different NT, L=4}. Furthermore, one should also note here that, contrary to the addition of more replicas, that extra computational effort would not be parallelizable in PA, since the annealing schedule must always be followed sequentially. %Therefore some disorder cases (actually majority) that are hard for PA algorithm can be sucessfully solved by DPA algorithm. Yet there are disorder cases which are hard for DPA algorithm that limiting it performance, and majority of them are not hard cases for PA. As hard disorder cases are actually limiting preformance of both algorithms, we observe interesting complementarity between both methods. The number of disorder cases that are hard for both algorithms is substantialy reduced, a feature that we will seek to explore and exploit in further study.

\subsubsection{$L = 6$}

\begin{figure}
\centering
 \includegraphics[width = \columnwidth]{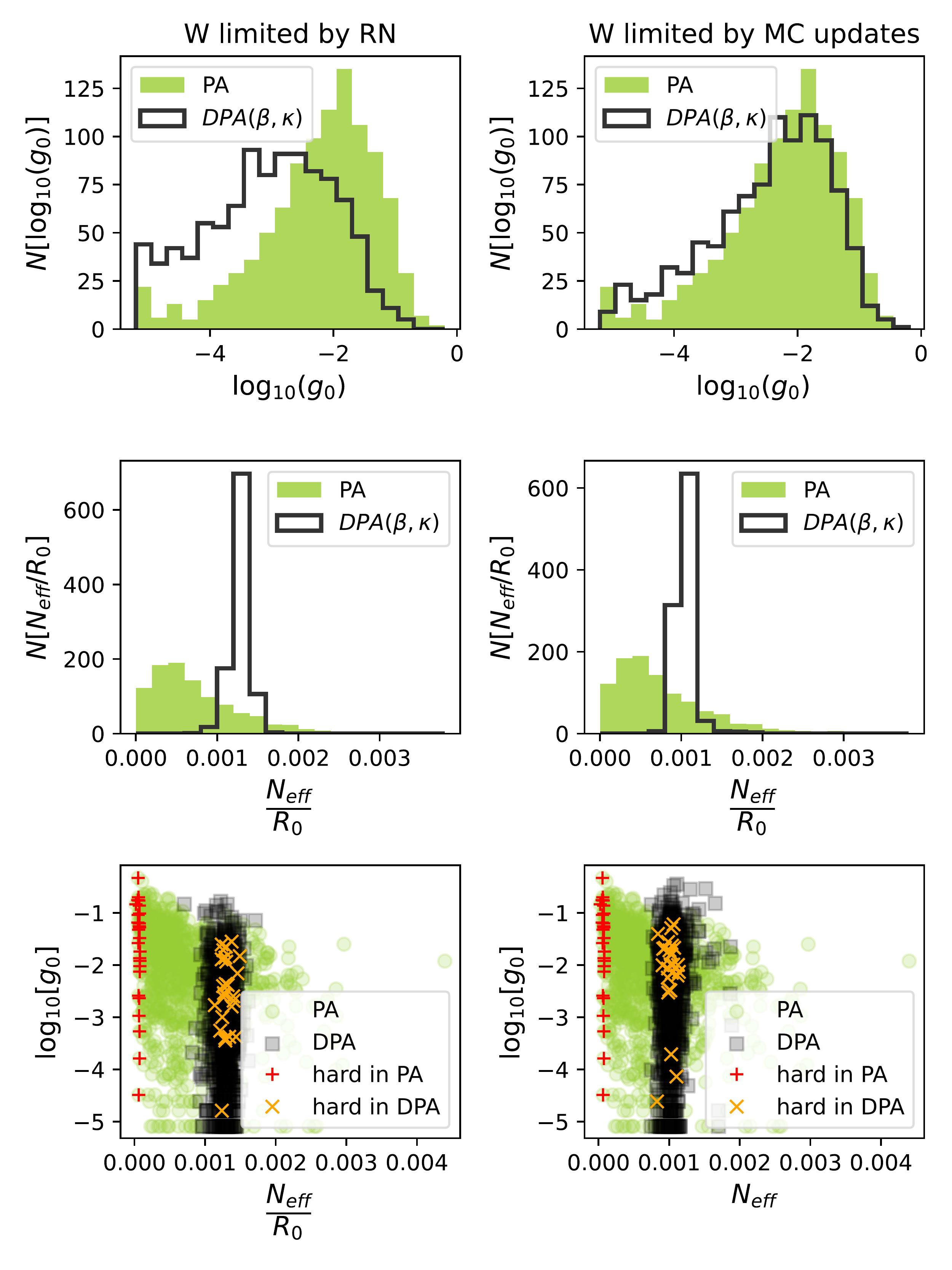}
 \caption{\textbf{DPA enables thermalization of $L=6$ lattices, even when these are hard for the PA algorithm.} Histograms of $g_0$ and $N_{\rm{eff}}/R$ obtained by solving, by both PA and DPA, a total of $N_{\rm{dis}} = 1000$ different disorders of $L=6$ RFW lattices, with bonds randomly distributed according to a normal distribution. Simulations using the same amount of random numbers (left) and of MCMC updates (right). Crosses in the bottom panels mark these instances PA finds the hardest (red), and how these same instances score in DPA (orange).}
  \label{fig: L6a}
\end{figure}

We now apply the same study to $L=6$ RFW lattices. As for the $L=4$ case, we solve several different Gaussian disorders with both PA and DPA considering the two scenarios discussed above, and plot the obtained histograms in Fig.~\ref{fig: L6a}. For the first scenario, (Fig.~\ref{fig: L6a}, left panels), we again see a certain trade-off between $g_0$ and $N_{\rm{eff}}$. Nevertheless, the gain in family entropy is now more noticeable than for smaller lattices, as now PA does encounter some hard instances for which it obtains a single surviving family, preventing adequate thermalization. On the contrary, the adaptive steps in DPA are capable of driving the population toward nonzero values, thus ensuring proper thermalization (Fig.~\ref{fig: L6a} center left panel). When considering the computational work limited by the number of MC updates, the $g_0$ obtained with DPA is comparable to that obtained with PA, while the gain in family entropy remains the same. Again, those hard instances of PA are properly thermalized under DPA. On top of that, comparing the results obtained for each individual disorder between the two scenarios (Fig.~\ref{fig: L6a} bottom panels), we see that, while for PA the hard instances are more or less evenly distributed along the range of $g_0$ values, they are shifted towards bigger values, especially for the second scenario.

\begin{figure}
\includegraphics[width=\columnwidth]{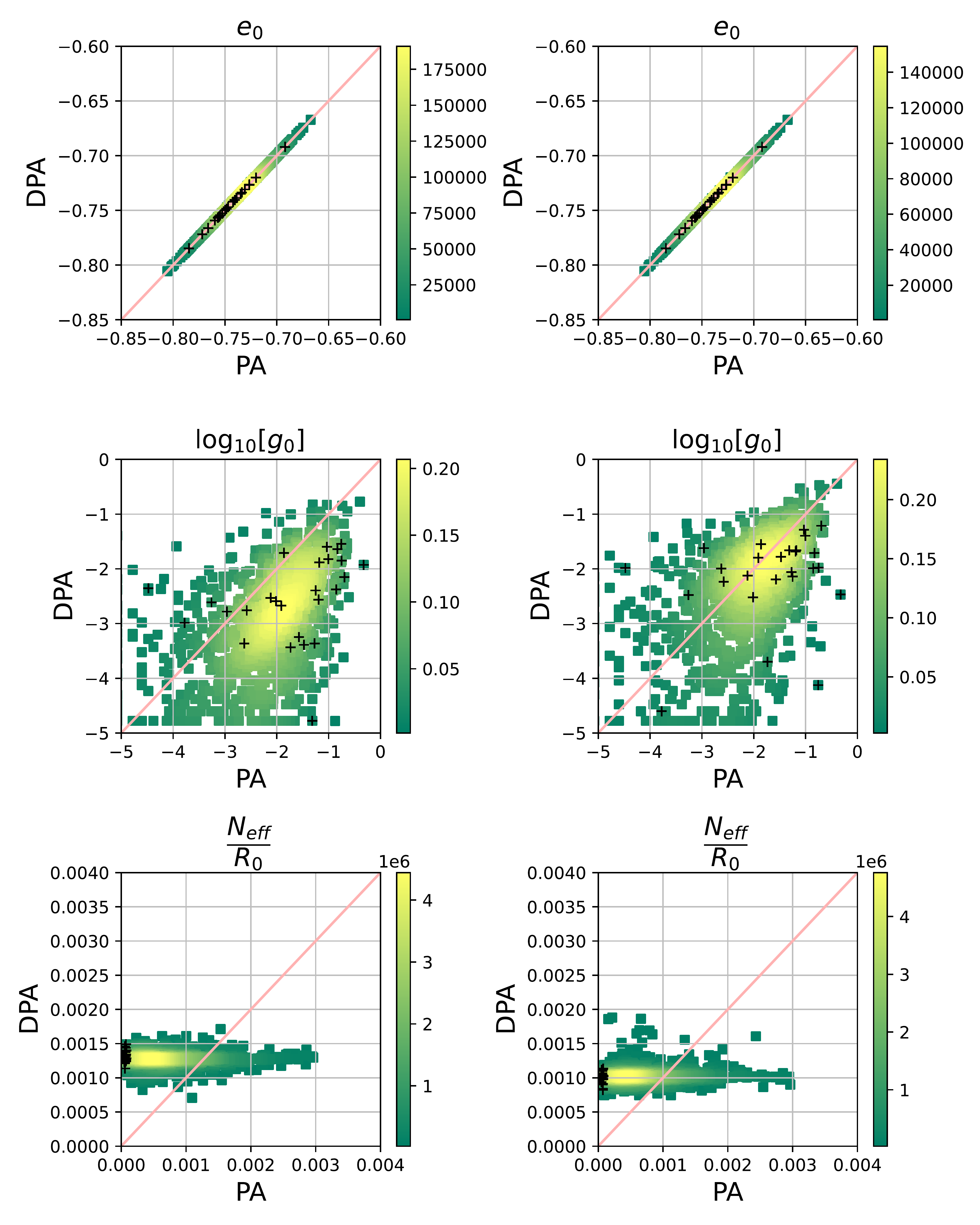}
 \caption{\textbf{DPA provides on average higher family entropies than PA when solving $L=6$ lattices.} Comparison of results obtained by PA and DPA for $L=6$ RFW lattices for the three measured parameters. From top to bottom: ground state energy, $g_0$ and effective number of surviving families. Simulations using the same amount of numbers (left) and of MCMC updates (right). Black crosses mark these instances that PA finds the hardest.}
  \label{fig: L6b}
\end{figure}

Looking at the top panels of Fig.~\ref{fig: L6b} we again confirm that both methods find the same ground-state energy for all disorder realizations. Speaking of $g_0$, the hardest instances are more or less uniformly distributed around their mean in both scenarios (Fig.~\ref{fig: L6b} middle panels), while shifted towards PA when restricted by the generation of random numbers, but centered between the two methods when restricted by the amount of MC updates.

\begin{figure}[ht]
\centering
 \includegraphics[scale = .23]{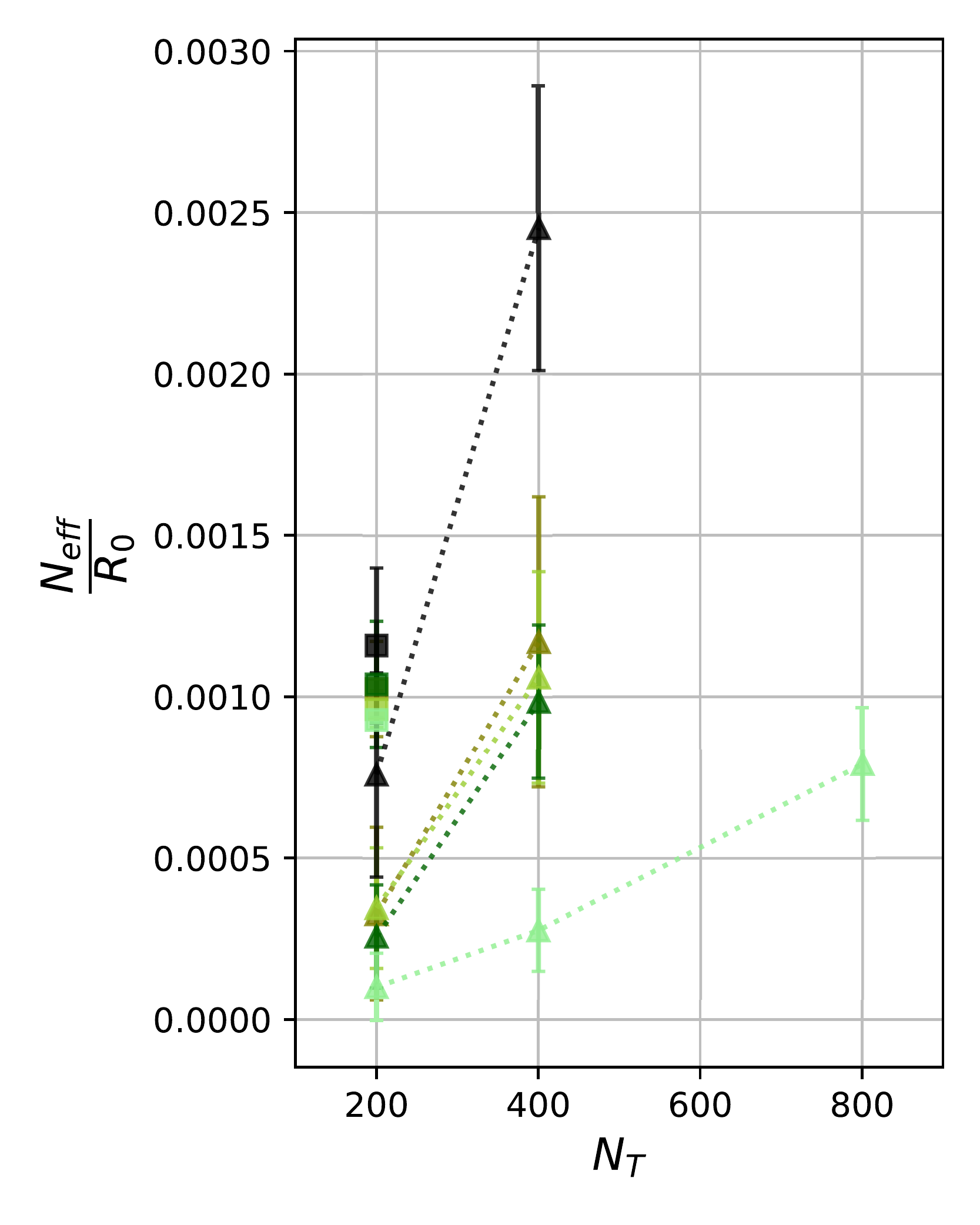}
 \caption{\textbf{Hard cases on PA require more adiabatic annealing processes to reach an equivaent family entropy to DPA.} Study of the amount of computational power that would have to be invested to solve $L=6$ RFW lattices with PA, in order to obtain a similar final family entropy as with DPA, for hard instances for which $g_0^{PA} \approx g_0^{DPA}$. Error bars are standard deviations over $N_{rep} = 100$ repetitions of the same process at each value of $N_T$. Square markers are obtained with DPA and triangle ones with PA using an increasing number of temperature steps. Each color indicates a different disorder.}
  \label{fig: different NT L=6}
\end{figure}

We finally study in Fig.~\ref{fig: different NT L=6}, as in the case $L=4$, a collection of instances that PA finds hard and that obtain a similar value of $g_0$ with both methods. Again, we use an increasingly adiabatic process in PA to see how much more computational power would be necessary to get results comparable to those obtained with PA. In this case, DPA achieves an equivalent performance to PA with about between 2 and 5 more computational investment, depending on the instance. We note a more linear behaviour of the family entropy with $N_T$ than in Fig.~\ref{fig: different NT L=6}, probably because the considered values of $N_T$ are not big enough. This indicates that $L=6$ is on the limit of our numerical capabilities.

\subsubsection{$L = 8$}
\begin{figure}[ht]
\centering
 \includegraphics[width = \columnwidth]{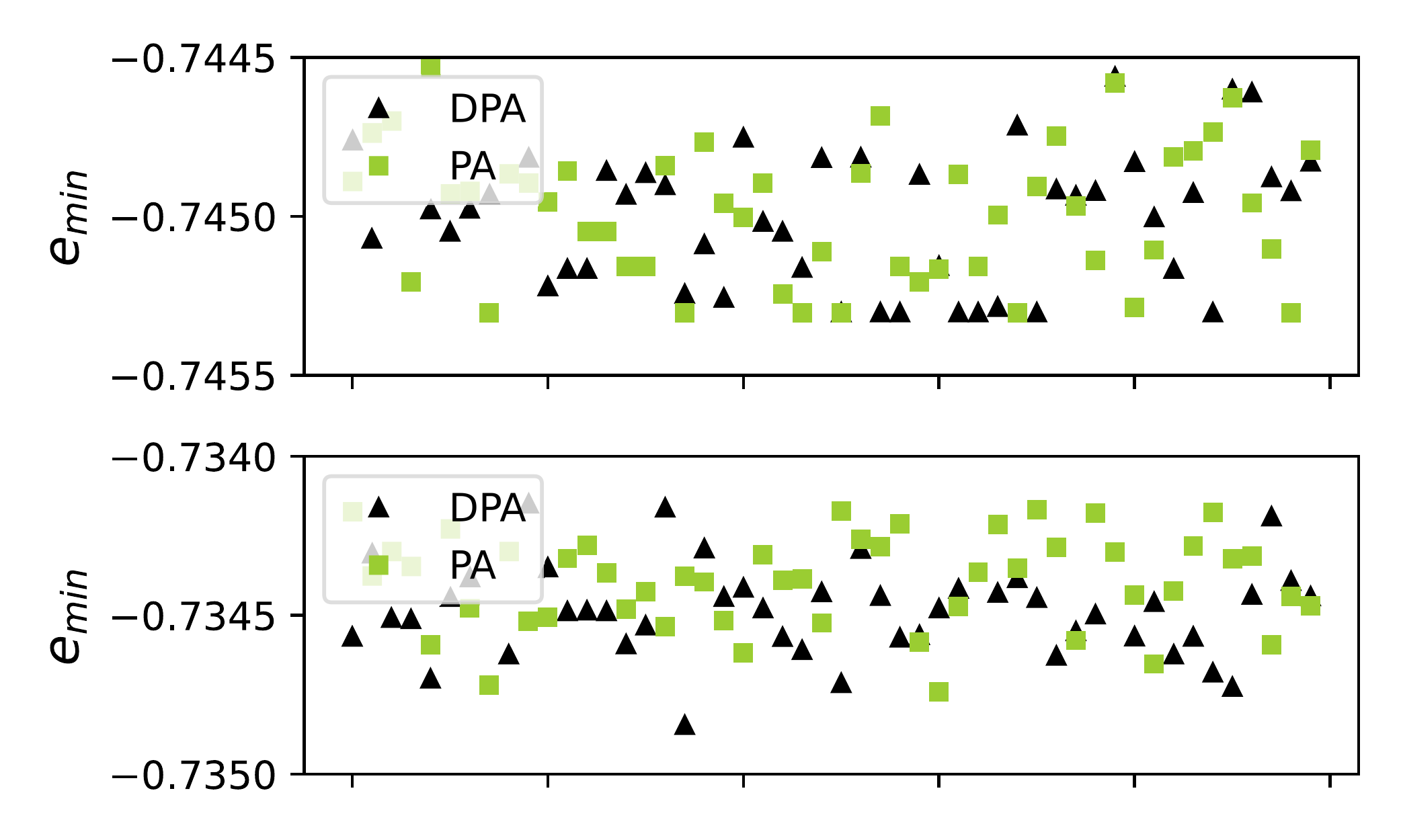}
 \caption{\textbf{Both algorithms have a low convergence ratio to the same minimum energy state among independent runs.} Upper panel: example of an easy case, for which both algorithms find the same solution after running it several times (both PA and DPA show a convergence ratio of $12\%$). Bottom panel: example of a hard case, for which both algorithms completely fail to converge towards any minimum energy state, and for which DPA finds the best solution between the two.}
  \label{fig: e0 spread L=8}
\end{figure}

In order to study hard EA disorders on big lattice sizes for which a proper thermalization is not guaranteed, one classically relies on running many times the same instance independently. We apply this methodology to $L=8$ RFW lattices with Gaussian disorder to test how both algorithms perform. Concretely, we study 20 different disorder instances and run each of them 50 times with each algorithm, considering the computational work limited by the amount of MC updates (second discussed scenario in previous sections). As the systems are not thermalized, we only focus here on the minimum energies per spin found during the simulations, $e_{min}$, and not on the previously discussed metrics. Contrarily to what should be expected from a thermalized system, none of the algorithms clearly converges to the same energy among different runs for a given disorder. In fact, when studying it with PA, for 11 out of the 20 studied cases the same minimum energy was not found in any of the runs. For 4 of them the minimum energy was found twice, for 2 of them four times and only for the 3 easiest ones it was repeatedly found six times.
In Fig.~\ref{fig: e0 spread L=8} we show, for one of the three easy studied instances (top panel) and for one of the 11 hard ones (bottom panel), the minimum energies found among independent runs by each algorithm.

\begin{figure}[ht]
\centering
 \includegraphics[width = \columnwidth]{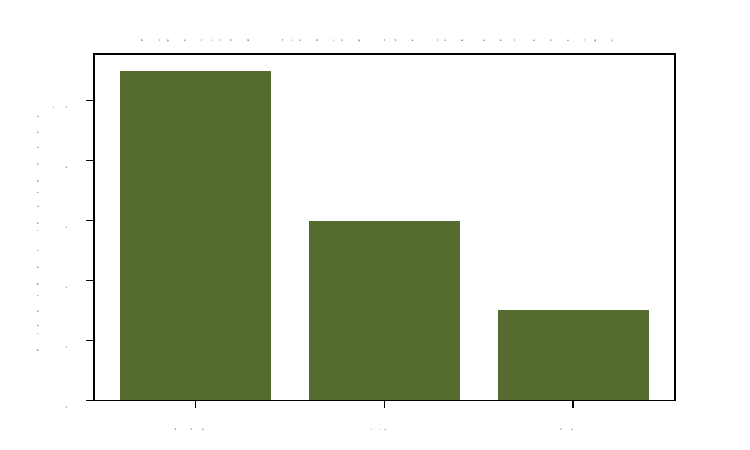}
 \caption{\textbf{DPA finds lower energy states.} When studying non-thermalized RFW $L=8$ lattices, DPA is generally able to find lower energies than PA using the same computational resources (in the same amount of MCMC updates scenario). Results obtained after solving 20 different disorder instances, 50 times each with each algorithm.}
  \label{fig: winning algorithm}
\end{figure}

On the one hand, we observe that the spread of the minimum energies found among several runs of the same disorder tends to be larger with DPA than with PA, probably due to the larger configurational space caused by the introduction of topological defects. On the other hand, when comparing the minimum energy found with both algorithms, we observe that DPA generally performs better, yielding lower energies (see Fig.~\ref{fig: winning algorithm}, the "tie" cases contain easy disorders). The fact that DPA generally finds states with lower energies when solving hard instances for which thermalization is poor (or, equivalently, for which no convergence to the same solution is achieved) indicates that the proposed non-local moves are indeed useful to explore energy landscape in hard disorder instances, effectively allowing the system escape local minima.

%We measure the spread of the minimum energies among different runs of the same instance by computing their median absolute deviations. In the right panel of Fig.~\ref{fig: e0 spread L=8} we show a box plot illustrating how these spreads are distributed among all the studied disorders for each of the algorithms. We observe that, generally, DPA yields a larger spread of minimum energies. On the other hand, when taking a look at the minimum energy found with both algorithms, we observe that DPA tends to perform better, Fig.~\ref{fig: winning algorithm}. While the larger spread of energies is probably due to the enlargement of the configurational space caused by the introduction of the topological defects, the fact that DPA generally finds states with lower energies indicates that the proposed non-local moves are indeed useful to overcome high energy barriers in hard instances.

\subsection{ Family entropy-preserving steps in PA}
\begin{figure}[ht]
\centering
 \includegraphics[width = \columnwidth]{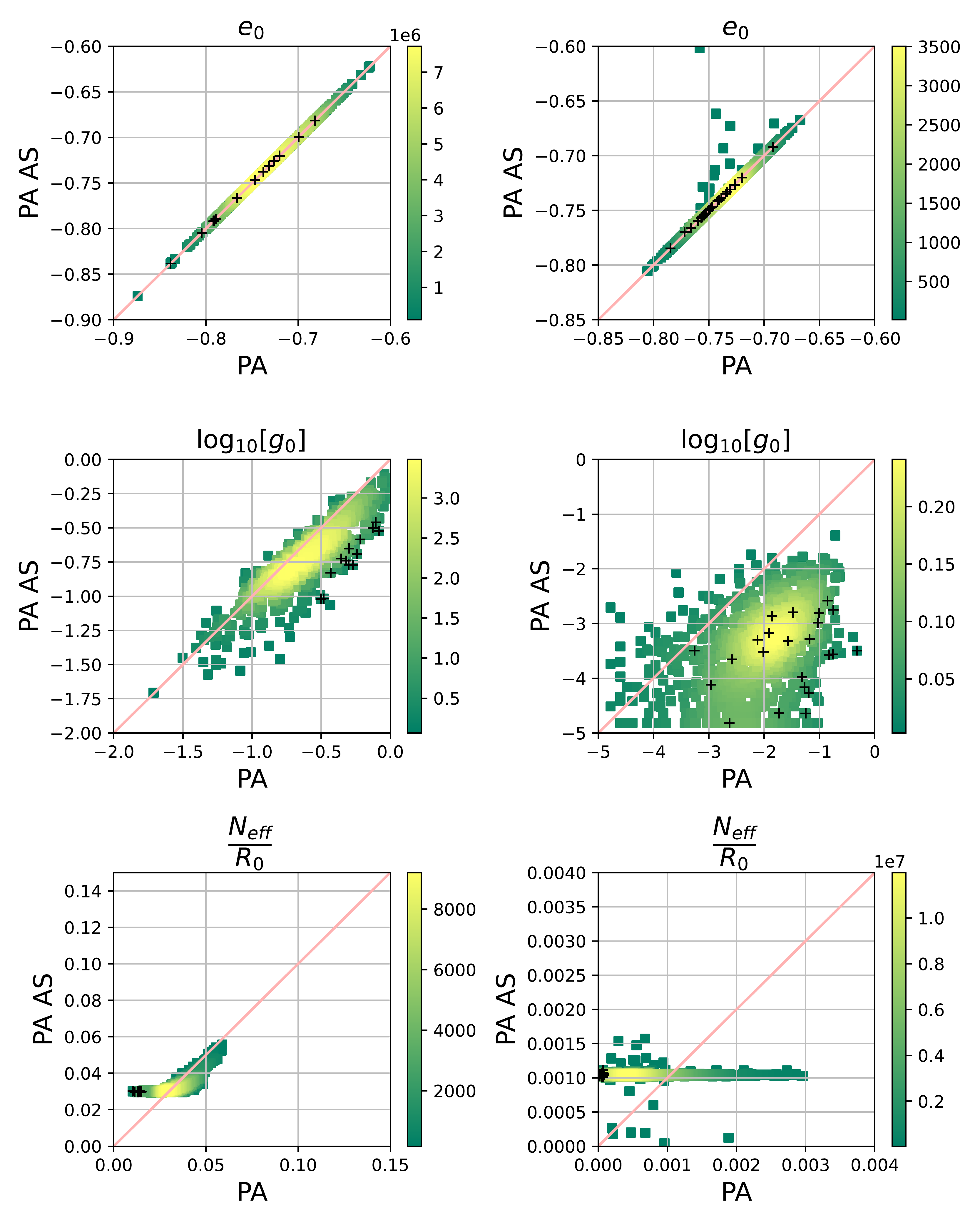}
 \caption{\textbf{The entropy-preserving steps are not optimal in standard PA.} For $L=4$ (left panels) and $L=6$ (right panels) RFW lattices the adaptive steps procedure is not effective when the parameter space is reduced to only temperature. Black crosses mark the disorders that normal PA finds the hardest to solve.}
  \label{fig: PA AS}
\end{figure}

To further test the effectivity of the newly introduced non-local moves, we implement the family entropy-preserving adaptives steps in PA. We use the same parameters used in the simulations of previous sections and find the results shown in Fig. \ref{fig: PA AS} for $L=4$ (left panels) and $L=6$ (right panels). While the adaptive steps are again able to generally impose a cut-off value on the final family entropy, the obtained $g_0$ is quite worse than in regular PA. For $L=6$ lattices the cut-off is actually not achieved for some disorders, some of which are even seen as easy (obtaining a high value of $N_{eff}$) by PA. This fact suggests that different annealing schedules may make PA find hard or easy different instances. More noticeably, the adaptive steps make PA fail to find the same ground state than when using the standard annealing schedule in some $L=6$ cases. We therefore conclude that the entropy-preserving annealing steps are not effective by themselves when restricted to a one dimensional annealing space, and thus the non-local moves must be the reason why DPA is able to improve thermalization of hard instances, as seen in previous sections.

\section{Conclusions and outlook} \label{sec: conclusions and outlook}
In this paper we introduced a method, inspired by the Toric Code, of exploring rugged energy landscape of 2d random-bond Ising and 3d random plaquette gauge models based on creation, movement and annihilation of topological defects. The advantages of this approach are: (i) from the point of view of the original phase space of the models, the effect of the dynamics of the topological defects is equivalent to non-local moves, which allow the system to overcome energy barriers that other exploration strategies based on thermal fluctuation or tunneling don't, (ii) topological defect moves are compatible with massively parallel implementations. The disadvantage of this approach is a substantial enlargement of the phase space in the early stages of simulation.

We implemented this method in a state-of-the-art Population Annealing algorithm by adding moves related to topological defects and an extra field parameter in the Hamiltonian, $\kappa$, which energetically penalizes them. The presented Defect-driven Population Annealing (DPA) algorithm utilizes annealing in a two-parameter space: the temperature and the $\kappa$, where starting temperature may now be arbitrary. This offers additional application flexibility, but leaves the annealing path undetermined. To this end we devised a new  family entropy-preserving adaptive step procedure that effectively navigates the $(\beta, \kappa)$-space in order to drive the replica population towards an objective value of final family entropy. The merging of more advanced annealing schemes such as culling fraction with the exploration of the two-dimensional annealing parameter space leaves room for improvement, which we leave as a possible extension of this work.

We have tested the new algorithm on the 3d random plaquette gauge model, as the 2d random bond Ising model lacks numerically hard disorder cases. We leave the problem of algorithm performance in 2d random bond Ising model for future study. The ground state problem of the 3d RPGM with Gaussian disorder was solved in a representation we named random field wall model, using standard PA as a benchmark. We found that the RFW model is substantially more numerically demanding than the 3d Edwards-Anderson model, which limited our thorough simulations to 
lattice sizes $L=4$ and $L=6$. Bigger lattices have been explored, but we leave its thorough study for a future extension of the present work. The results for DPA on those lattices have shown that it is capable of improving thermalization in comparison with PA, effectively avoiding running out of family entropy even for the hardest disorder instances. 

When focusing on the instances that PA finds the hardest, we have been able to observe that DPA can be superior to regular PA with a computational investment 2 to 16 times higher, approximately, depending on the case. Yet sometimes DPA trades improvement of family entropy for decrease in $g_0$ metric. Finally, when studying hard cases for which thermalization is not properly achieved, the proposed non-local moves still show an advantage in overcoming energy barriers and thus yield lower energies.
It is also worth noting that the results are greatly improved when a fast enough random number generator is used, such that the Monte Carlo updates themselves constitute the real bottleneck for the simulations and thus the fair measure of the computational work invested.

\section{Acknowledgments}
D.C. and P.R.G. would like to thank H. Katzgraber and J. Machta for illuminating discussion and comments that helped to improve manuscript.
D.C. acknowledges funding from Generalitat de Catalunya (AGAUR Doctorats Industrials 2019, 2n termini).
ICFO group acknowledges support from: ERC AdG NOQIA; Ministerio de Ciencia y Innovation Agencia Estatal de Investigaciones (PGC2018-097027-B-I00/10.13039/501100011033,  CEX2019-000910-S/10.13039/501100011033, Plan National FIDEUA PID2019-106901GB-I00, FPI, QUANTERA MAQS PCI2019-111828-2, QUANTERA DYNAMITE PCI2022-132919,  Proyectos de I+D+I “Retos Colaboración” QUSPIN RTC2019-007196-7); MICIIN with funding from European Union NextGenerationEU (PRTR-C17.I1) and by Generalitat de Catalunya;  Fundació Cellex; Fundació Mir-Puig; Generalitat de Catalunya (European Social Fund FEDER and CERCA program, AGAUR Grant No. 2021 SGR 01452, QuantumCAT \ U16-011424, co-funded by ERDF Operational Program of Catalonia 2014-2020); Barcelona Supercomputing Center MareNostrum (FI-2023-1-0013); EU (PASQuanS2.1, 101113690); EU Horizon 2020 FET-OPEN OPTOlogic (Grant No 899794); EU Horizon Europe Program (Grant Agreement 101080086 — NeQST), National Science Centre, Poland (Symfonia Grant No. 2016/20/W/ST4/00314); ICFO Internal “QuantumGaudi” project; European Union’s Horizon 2020 research and innovation program under the Marie-Skłodowska-Curie grant agreement No 101029393 (STREDCH) and No 847648  (“La Caixa” Junior Leaders fellowships ID100010434: LCF/BQ/PI19/11690013, LCF/BQ/PI20/11760031,  LCF/BQ/PR20/11770012, LCF/BQ/PR21/11840013). Views and opinions expressed are, however, those of the author(s) only and do not necessarily reflect those of the European Union, European Commission, European Climate, Infrastructure and Environment Executive Agency (CINEA), nor any other granting authority.  Neither the European Union nor any granting authority can be held responsible for them. M.Á.G.-M. acknowledges funding from the Spanish Ministry of Education and Professional Training (MEFP) through the Beatriz Galindo program 2018 (BEAGAL18/00203), QuantERA II Cofund 2021 PCI2022-133004, Projects of MCIN with funding from European Union NextGenerationEU (PRTR-C17.I1) and by Generalitat Valenciana, with Ref. 20220883 (PerovsQuTe) and COMCUANTICA/007 (QuanTwin), and Red Temática RED2022-134391-T.

\bibliography{biblio}

%apsrev4-2.bst 2019-01-14 (MD) hand-edited version of apsrev4-1.bst
%Control: key (0)
%Control: author (8) initials jnrlst
%Control: editor formatted (1) identically to author
%Control: production of article title (0) allowed
%Control: page (0) single
%Control: year (1) truncated
%Control: production of eprint (0) enabled
\begin{thebibliography}{41}%
\makeatletter
\providecommand \@ifxundefined [1]{%
 \@ifx{#1\undefined}
}%
\providecommand \@ifnum [1]{%
 \ifnum #1\expandafter \@firstoftwo
 \else \expandafter \@secondoftwo
 \fi
}%
\providecommand \@ifx [1]{%
 \ifx #1\expandafter \@firstoftwo
 \else \expandafter \@secondoftwo
 \fi
}%
\providecommand \natexlab [1]{#1}%
\providecommand \enquote  [1]{``#1''}%
\providecommand \bibnamefont  [1]{#1}%
\providecommand \bibfnamefont [1]{#1}%
\providecommand \citenamefont [1]{#1}%
\providecommand \href@noop [0]{\@secondoftwo}%
\providecommand \href [0]{\begingroup \@sanitize@url \@href}%
\providecommand \@href[1]{\@@startlink{#1}\@@href}%
\providecommand \@@href[1]{\endgroup#1\@@endlink}%
\providecommand \@sanitize@url [0]{\catcode `\\12\catcode `\$12\catcode `\&12\catcode `\#12\catcode `\^12\catcode `\_12\catcode `\%12\relax}%
\providecommand \@@startlink[1]{}%
\providecommand \@@endlink[0]{}%
\providecommand \url  [0]{\begingroup\@sanitize@url \@url }%
\providecommand \@url [1]{\endgroup\@href {#1}{\urlprefix }}%
\providecommand \urlprefix  [0]{URL }%
\providecommand \Eprint [0]{\href }%
\providecommand \doibase [0]{https://doi.org/}%
\providecommand \selectlanguage [0]{\@gobble}%
\providecommand \bibinfo  [0]{\@secondoftwo}%
\providecommand \bibfield  [0]{\@secondoftwo}%
\providecommand \translation [1]{[#1]}%
\providecommand \BibitemOpen [0]{}%
\providecommand \bibitemStop [0]{}%
\providecommand \bibitemNoStop [0]{.\EOS\space}%
\providecommand \EOS [0]{\spacefactor3000\relax}%
\providecommand \BibitemShut  [1]{\csname bibitem#1\endcsname}%
\let\auto@bib@innerbib\@empty
%</preamble>
\bibitem [{\citenamefont {Brush}(1967)}]{brush}%
  \BibitemOpen
  \bibfield  {author} {\bibinfo {author} {\bibfnamefont {S.}~\bibnamefont {Brush}},\ }\bibfield  {title} {\bibinfo {title} {History of the lenz-ising model},\ }\href@noop {} {\bibfield  {journal} {\bibinfo  {journal} {Rev. Mod. Phys.}\ }\textbf {\bibinfo {volume} {39}},\ \bibinfo {pages} {883} (\bibinfo {year} {1967})}\BibitemShut {NoStop}%
\bibitem [{\citenamefont {Parisi}(2007)}]{parisi2007mean}%
  \BibitemOpen
  \bibfield  {author} {\bibinfo {author} {\bibfnamefont {G.}~\bibnamefont {Parisi}},\ }\href@noop {} {\bibinfo {title} {Mean field theory of spin glasses: statics and dynamics}} (\bibinfo {year} {2007}),\ \Eprint {https://arxiv.org/abs/0706.0094} {arXiv:0706.0094 [cond-mat.dis-nn]} \BibitemShut {NoStop}%
\bibitem [{\citenamefont {Barahona}(1982)}]{Barahona1982}%
  \BibitemOpen
  \bibfield  {author} {\bibinfo {author} {\bibfnamefont {F.}~\bibnamefont {Barahona}},\ }\bibfield  {title} {\bibinfo {title} {On the computational complexity of ising spin glass models},\ }\href@noop {} {\bibfield  {journal} {\bibinfo  {journal} {Journal of Physics A: Mathematical and General}\ }\textbf {\bibinfo {volume} {15}},\ \bibinfo {pages} {3241} (\bibinfo {year} {1982})}\BibitemShut {NoStop}%
\bibitem [{\citenamefont {Amit}\ \emph {et~al.}(1985)\citenamefont {Amit}, \citenamefont {Gutfreund},\ and\ \citenamefont {Sompolinsky}}]{Amit1985}%
  \BibitemOpen
  \bibfield  {author} {\bibinfo {author} {\bibfnamefont {D.}~\bibnamefont {Amit}}, \bibinfo {author} {\bibfnamefont {H.}~\bibnamefont {Gutfreund}},\ and\ \bibinfo {author} {\bibfnamefont {H.}~\bibnamefont {Sompolinsky}},\ }\bibfield  {title} {\bibinfo {title} {Spin-glass models of neural networks},\ }\href@noop {} {\bibfield  {journal} {\bibinfo  {journal} {Phys. Rev. A}\ }\textbf {\bibinfo {volume} {32}},\ \bibinfo {pages} {1007} (\bibinfo {year} {1985})}\BibitemShut {NoStop}%
\bibitem [{\citenamefont {van Hemmen}(1986)}]{vanHemmen1986}%
  \BibitemOpen
  \bibfield  {author} {\bibinfo {author} {\bibfnamefont {J.}~\bibnamefont {van Hemmen}},\ }\bibfield  {title} {\bibinfo {title} {Spin-glass models of a neural network},\ }\href@noop {} {\bibfield  {journal} {\bibinfo  {journal} {Phys. Rev. A}\ }\textbf {\bibinfo {volume} {34}},\ \bibinfo {pages} {3435} (\bibinfo {year} {1986})}\BibitemShut {NoStop}%
\bibitem [{\citenamefont {Lucas}(2014)}]{Lucas2014}%
  \BibitemOpen
  \bibfield  {author} {\bibinfo {author} {\bibfnamefont {A.}~\bibnamefont {Lucas}},\ }\bibfield  {title} {\bibinfo {title} {Ising formulations of many np problems},\ }\href@noop {} {\bibfield  {journal} {\bibinfo  {journal} {Frontiers in Physics}\ }\textbf {\bibinfo {volume} {2}} (\bibinfo {year} {2014})}\BibitemShut {NoStop}%
\bibitem [{\citenamefont {Lucas}(2019)}]{Lucas2019}%
  \BibitemOpen
  \bibfield  {author} {\bibinfo {author} {\bibfnamefont {A.}~\bibnamefont {Lucas}},\ }\bibfield  {title} {\bibinfo {title} {Hard combinatorial problems and minor embeddings on lattice graphs},\ }\href@noop {} {\bibfield  {journal} {\bibinfo  {journal} {Quantum Information Processing}\ }\textbf {\bibinfo {volume} {18}} (\bibinfo {year} {2019})}\BibitemShut {NoStop}%
\bibitem [{\citenamefont {Irb\"ack}\ \emph {et~al.}(2022)\citenamefont {Irb\"ack}, \citenamefont {Knuthson}, \citenamefont {Mohanty},\ and\ \citenamefont {Peterson}}]{Irback2022}%
  \BibitemOpen
  \bibfield  {author} {\bibinfo {author} {\bibfnamefont {A.}~\bibnamefont {Irb\"ack}}, \bibinfo {author} {\bibfnamefont {L.}~\bibnamefont {Knuthson}}, \bibinfo {author} {\bibfnamefont {S.}~\bibnamefont {Mohanty}},\ and\ \bibinfo {author} {\bibfnamefont {C.}~\bibnamefont {Peterson}},\ }\bibfield  {title} {\bibinfo {title} {Folding lattice proteins with quantum annealing},\ }\href@noop {} {\bibfield  {journal} {\bibinfo  {journal} {Phys. Rev. Res.}\ }\textbf {\bibinfo {volume} {4}},\ \bibinfo {pages} {043013} (\bibinfo {year} {2022})}\BibitemShut {NoStop}%
\bibitem [{\citenamefont {Oliveira}\ \emph {et~al.}(2018)\citenamefont {Oliveira}, \citenamefont {Silva},\ and\ \citenamefont {Oliveira}}]{Oliveira2018}%
  \BibitemOpen
  \bibfield  {author} {\bibinfo {author} {\bibfnamefont {N.~D.}\ \bibnamefont {Oliveira}}, \bibinfo {author} {\bibfnamefont {R.~D.~A.}\ \bibnamefont {Silva}},\ and\ \bibinfo {author} {\bibfnamefont {W.~D.}\ \bibnamefont {Oliveira}},\ }\bibfield  {title} {\bibinfo {title} {Qubo formulation for the contact map overlap problem},\ }\href@noop {} {\bibfield  {journal} {\bibinfo  {journal} {International Journal of Quantum Information}\ }\textbf {\bibinfo {volume} {16}},\ \bibinfo {pages} {1840007} (\bibinfo {year} {2018})}\BibitemShut {NoStop}%
\bibitem [{\citenamefont {Hernández}\ \emph {et~al.}(2020)\citenamefont {Hernández}, \citenamefont {Díaz}, \citenamefont {Forets},\ and\ \citenamefont {Sotelo}}]{Hernandez2020}%
  \BibitemOpen
  \bibfield  {author} {\bibinfo {author} {\bibfnamefont {F.}~\bibnamefont {Hernández}}, \bibinfo {author} {\bibfnamefont {K.}~\bibnamefont {Díaz}}, \bibinfo {author} {\bibfnamefont {M.}~\bibnamefont {Forets}},\ and\ \bibinfo {author} {\bibfnamefont {R.}~\bibnamefont {Sotelo}},\ }\bibfield  {title} {\bibinfo {title} {Application of quantum optimization techniques (qubo method) to cargo logistics on ships and airplanes}\ }(\bibinfo {year} {2020})\ pp.\ \bibinfo {pages} {1--1}\BibitemShut {NoStop}%
\bibitem [{\citenamefont {Phillipson}\ and\ \citenamefont {Chiscop}(2021)}]{Phillipson2021}%
  \BibitemOpen
  \bibfield  {author} {\bibinfo {author} {\bibfnamefont {F.}~\bibnamefont {Phillipson}}\ and\ \bibinfo {author} {\bibfnamefont {I.}~\bibnamefont {Chiscop}},\ }\bibfield  {title} {\bibinfo {title} {Multimodal container planning: A qubo formulation and implementation on a quantum annealer}\ }(\bibinfo  {publisher} {Springer International Publishing},\ \bibinfo {address} {Cham},\ \bibinfo {year} {2021})\BibitemShut {NoStop}%
\bibitem [{\citenamefont {Ariño}\ and\ \citenamefont {Palacios}(2023)}]{sales2023}%
  \BibitemOpen
  \bibfield  {author} {\bibinfo {author} {\bibfnamefont {J.}~\bibnamefont {Ariño}}\ and\ \bibinfo {author} {\bibfnamefont {R.}~\bibnamefont {Palacios}},\ }\href@noop {} {\bibinfo {title} {Adiabatic quantum computing for logistic transport optimization}} (\bibinfo {year} {2023}),\ \Eprint {https://arxiv.org/abs/2301.07691} {arXiv:2301.07691 [quant-ph]} \BibitemShut {NoStop}%
\bibitem [{\citenamefont {Kirkpatrick}\ \emph {et~al.}(1983)\citenamefont {Kirkpatrick}, \citenamefont {Gelatt},\ and\ \citenamefont {Vecchi}}]{Kirkpatrick1983}%
  \BibitemOpen
  \bibfield  {author} {\bibinfo {author} {\bibfnamefont {S.}~\bibnamefont {Kirkpatrick}}, \bibinfo {author} {\bibfnamefont {C.}~\bibnamefont {Gelatt}},\ and\ \bibinfo {author} {\bibfnamefont {M.}~\bibnamefont {Vecchi}},\ }\bibfield  {title} {\bibinfo {title} {Optimization by simulated annealing},\ }\href@noop {} {\bibfield  {journal} {\bibinfo  {journal} {Science}\ }\textbf {\bibinfo {volume} {220}},\ \bibinfo {pages} {671} (\bibinfo {year} {1983})}\BibitemShut {NoStop}%
\bibitem [{\citenamefont {Gubernatis}(2003)}]{Gubernatis2003}%
  \BibitemOpen
  \bibfield  {author} {\bibinfo {author} {\bibfnamefont {J.}~\bibnamefont {Gubernatis}},\ }\href@noop {} {\emph {\bibinfo {title} {The Monte Carlo Method in the Physical Sciences: Celebrating the 50th Anniversary of the Metropolis Algorithm}}}\ (\bibinfo  {publisher} {American Institute of Physics},\ \bibinfo {year} {2003})\BibitemShut {NoStop}%
\bibitem [{\citenamefont {Machta}(2010)}]{Machta2010}%
  \BibitemOpen
  \bibfield  {author} {\bibinfo {author} {\bibfnamefont {J.}~\bibnamefont {Machta}},\ }\bibfield  {title} {\bibinfo {title} {Population annealing with weighted averages: A monte carlo method for rough free-energy landscapes},\ }\href@noop {} {\bibfield  {journal} {\bibinfo  {journal} {Phys. Rev. E}\ }\textbf {\bibinfo {volume} {82}},\ \bibinfo {pages} {026704} (\bibinfo {year} {2010})}\BibitemShut {NoStop}%
\bibitem [{\citenamefont {Wang}\ \emph {et~al.}(2015{\natexlab{a}})\citenamefont {Wang}, \citenamefont {Machta},\ and\ \citenamefont {Katzgraber}}]{Wang2015}%
  \BibitemOpen
  \bibfield  {author} {\bibinfo {author} {\bibfnamefont {W.}~\bibnamefont {Wang}}, \bibinfo {author} {\bibfnamefont {J.}~\bibnamefont {Machta}},\ and\ \bibinfo {author} {\bibfnamefont {H.}~\bibnamefont {Katzgraber}},\ }\bibfield  {title} {\bibinfo {title} {Comparing monte carlo methods for finding ground states of ising spin glasses: Population annealing, simulated annealing, and parallel tempering},\ }\href@noop {} {\bibfield  {journal} {\bibinfo  {journal} {Phys. Rev. E}\ }\textbf {\bibinfo {volume} {92}},\ \bibinfo {pages} {013303} (\bibinfo {year} {2015}{\natexlab{a}})}\BibitemShut {NoStop}%
\bibitem [{\citenamefont {Wang}\ \emph {et~al.}(2015{\natexlab{b}})\citenamefont {Wang}, \citenamefont {Machta},\ and\ \citenamefont {Katzgraber}}]{Wan2015B}%
  \BibitemOpen
  \bibfield  {author} {\bibinfo {author} {\bibfnamefont {W.}~\bibnamefont {Wang}}, \bibinfo {author} {\bibfnamefont {J.}~\bibnamefont {Machta}},\ and\ \bibinfo {author} {\bibfnamefont {H.}~\bibnamefont {Katzgraber}},\ }\bibfield  {title} {\bibinfo {title} {Population annealing: Theory and application in spin glasses},\ }\href@noop {} {\bibfield  {journal} {\bibinfo  {journal} {Phys. Rev. E}\ }\textbf {\bibinfo {volume} {92}},\ \bibinfo {pages} {063307} (\bibinfo {year} {2015}{\natexlab{b}})}\BibitemShut {NoStop}%
\bibitem [{\citenamefont {Metropolis}(1953)}]{Metropolis1953}%
  \BibitemOpen
  \bibfield  {author} {\bibinfo {author} {\bibfnamefont {N.~e.~a.}\ \bibnamefont {Metropolis}},\ }\bibfield  {title} {\bibinfo {title} {Equations of state calculations by fast computing machines},\ }\href@noop {} {\bibfield  {journal} {\bibinfo  {journal} {The Journal of Chemical Physics}\ }\textbf {\bibinfo {volume} {21}} (\bibinfo {year} {1953})}\BibitemShut {NoStop}%
\bibitem [{\citenamefont {Newman}\ and\ \citenamefont {Barkema}(1999)}]{newmanb99}%
  \BibitemOpen
  \bibfield  {author} {\bibinfo {author} {\bibfnamefont {M.}~\bibnamefont {Newman}}\ and\ \bibinfo {author} {\bibfnamefont {G.}~\bibnamefont {Barkema}},\ }\href@noop {} {\emph {\bibinfo {title} {Monte Carlo methods in statistical physics}}}\ (\bibinfo  {publisher} {Oxford university press},\ \bibinfo {year} {1999})\BibitemShut {NoStop}%
\bibitem [{\citenamefont {Edwards}\ and\ \citenamefont {Anderson}(1975)}]{EdwardsAnderson1975}%
  \BibitemOpen
  \bibfield  {author} {\bibinfo {author} {\bibfnamefont {S.}~\bibnamefont {Edwards}}\ and\ \bibinfo {author} {\bibfnamefont {P.}~\bibnamefont {Anderson}},\ }\bibfield  {title} {\bibinfo {title} {Theory of spin glasses},\ }\href@noop {} {\bibfield  {journal} {\bibinfo  {journal} {Journal of Physics F: Metal Physics}\ }\textbf {\bibinfo {volume} {5}},\ \bibinfo {pages} {965} (\bibinfo {year} {1975})}\BibitemShut {NoStop}%
\bibitem [{\citenamefont {Swendsen}\ and\ \citenamefont {Wang}(1987)}]{PhysRevLett.58.86}%
  \BibitemOpen
  \bibfield  {author} {\bibinfo {author} {\bibfnamefont {R.}~\bibnamefont {Swendsen}}\ and\ \bibinfo {author} {\bibfnamefont {J.}~\bibnamefont {Wang}},\ }\bibfield  {title} {\bibinfo {title} {Nonuniversal critical dynamics in monte carlo simulations},\ }\href@noop {} {\bibfield  {journal} {\bibinfo  {journal} {Phys. Rev. Lett.}\ }\textbf {\bibinfo {volume} {58}},\ \bibinfo {pages} {86} (\bibinfo {year} {1987})}\BibitemShut {NoStop}%
\bibitem [{\citenamefont {Wolff}(1989)}]{Wolff1989}%
  \BibitemOpen
  \bibfield  {author} {\bibinfo {author} {\bibfnamefont {U.}~\bibnamefont {Wolff}},\ }\bibfield  {title} {\bibinfo {title} {Collective monte carlo updating for spin systems},\ }\href@noop {} {\bibfield  {journal} {\bibinfo  {journal} {Phys. Rev. Lett.}\ }\textbf {\bibinfo {volume} {62}},\ \bibinfo {pages} {361} (\bibinfo {year} {1989})}\BibitemShut {NoStop}%
\bibitem [{\citenamefont {Houdayer}(2001)}]{Houdayer2001}%
  \BibitemOpen
  \bibfield  {author} {\bibinfo {author} {\bibfnamefont {J.}~\bibnamefont {Houdayer}},\ }\bibfield  {title} {\bibinfo {title} {A cluster monte carlo algorithm for 2-dimensional spin glasses},\ }\href@noop {} {\bibfield  {journal} {\bibinfo  {journal} {The European Physical Journal B - Condensed Matter and Complex Systems}\ }\textbf {\bibinfo {volume} {22}},\ \bibinfo {pages} {479} (\bibinfo {year} {2001})}\BibitemShut {NoStop}%
\bibitem [{\citenamefont {Zhu}\ \emph {et~al.}(2015)\citenamefont {Zhu}, \citenamefont {Ochoa},\ and\ \citenamefont {Katzgraber}}]{PhysRevLett.115.077201}%
  \BibitemOpen
  \bibfield  {author} {\bibinfo {author} {\bibfnamefont {Z.}~\bibnamefont {Zhu}}, \bibinfo {author} {\bibfnamefont {A.}~\bibnamefont {Ochoa}},\ and\ \bibinfo {author} {\bibfnamefont {H.}~\bibnamefont {Katzgraber}},\ }\bibfield  {title} {\bibinfo {title} {Efficient cluster algorithm for spin glasses in any space dimension},\ }\href@noop {} {\bibfield  {journal} {\bibinfo  {journal} {Phys. Rev. Lett.}\ }\textbf {\bibinfo {volume} {115}},\ \bibinfo {pages} {077201} (\bibinfo {year} {2015})}\BibitemShut {NoStop}%
\bibitem [{\citenamefont {Finnila}\ \emph {et~al.}(1994)\citenamefont {Finnila}, \citenamefont {Gomez}, \citenamefont {Sebenik}, \citenamefont {Stenson},\ and\ \citenamefont {Doll}}]{FINNILA19943}%
  \BibitemOpen
  \bibfield  {author} {\bibinfo {author} {\bibfnamefont {A.}~\bibnamefont {Finnila}}, \bibinfo {author} {\bibfnamefont {M.}~\bibnamefont {Gomez}}, \bibinfo {author} {\bibfnamefont {C.}~\bibnamefont {Sebenik}}, \bibinfo {author} {\bibfnamefont {C.}~\bibnamefont {Stenson}},\ and\ \bibinfo {author} {\bibfnamefont {J.}~\bibnamefont {Doll}},\ }\bibfield  {title} {\bibinfo {title} {Quantum annealing: A new method for minimizing multidimensional functions},\ }\href@noop {} {\bibfield  {journal} {\bibinfo  {journal} {Chemical Physics Letters}\ }\textbf {\bibinfo {volume} {219}},\ \bibinfo {pages} {343} (\bibinfo {year} {1994})}\BibitemShut {NoStop}%
\bibitem [{\citenamefont {Morita}(2008)}]{Satoshi2008}%
  \BibitemOpen
  \bibfield  {author} {\bibinfo {author} {\bibfnamefont {S.}~\bibnamefont {Morita}},\ }\bibfield  {title} {\bibinfo {title} {Mathematical foundation of quantum annealing},\ }\href@noop {} {\bibfield  {journal} {\bibinfo  {journal} {Journal of Mathematical Physics}\ }\textbf {\bibinfo {volume} {49}} (\bibinfo {year} {2008})}\BibitemShut {NoStop}%
\bibitem [{\citenamefont {Crosson}\ and\ \citenamefont {Harrow}(2016)}]{Crosson2016}%
  \BibitemOpen
  \bibfield  {author} {\bibinfo {author} {\bibfnamefont {E.}~\bibnamefont {Crosson}}\ and\ \bibinfo {author} {\bibfnamefont {A.~W.}\ \bibnamefont {Harrow}},\ }\bibfield  {title} {\bibinfo {title} {Simulated quantum annealing can be exponentially faster than classical simulated annealing},\ }in\ \href@noop {} {\emph {\bibinfo {booktitle} {2016 IEEE 57th Annual Symposium on Foundations of Computer Science (FOCS)}}}\ (\bibinfo {organization} {IEEE},\ \bibinfo {year} {2016})\ pp.\ \bibinfo {pages} {714--723}\BibitemShut {NoStop}%
\bibitem [{\citenamefont {Kitaev}(2003)}]{KITAEV2003}%
  \BibitemOpen
  \bibfield  {author} {\bibinfo {author} {\bibfnamefont {A.}~\bibnamefont {Kitaev}},\ }\bibfield  {title} {\bibinfo {title} {Fault-tolerant quantum computation by anyons},\ }\href@noop {} {\bibfield  {journal} {\bibinfo  {journal} {Annals of Physics}\ }\textbf {\bibinfo {volume} {303}},\ \bibinfo {pages} {2} (\bibinfo {year} {2003})}\BibitemShut {NoStop}%
\bibitem [{\citenamefont {Dennis}\ \emph {et~al.}(2002)\citenamefont {Dennis}, \citenamefont {Kitaev}, \citenamefont {Landahl},\ and\ \citenamefont {Preskill}}]{10.1063/1.1499754}%
  \BibitemOpen
  \bibfield  {author} {\bibinfo {author} {\bibfnamefont {E.}~\bibnamefont {Dennis}}, \bibinfo {author} {\bibfnamefont {A.}~\bibnamefont {Kitaev}}, \bibinfo {author} {\bibfnamefont {A.}~\bibnamefont {Landahl}},\ and\ \bibinfo {author} {\bibfnamefont {J.}~\bibnamefont {Preskill}},\ }\bibfield  {title} {\bibinfo {title} {Topological quantum memory},\ }\href@noop {} {\bibfield  {journal} {\bibinfo  {journal} {Journal of Mathematical Physics}\ }\textbf {\bibinfo {volume} {43}},\ \bibinfo {pages} {4452} (\bibinfo {year} {2002})}\BibitemShut {NoStop}%
\bibitem [{\citenamefont {Palassini}\ and\ \citenamefont {Young}(1999)}]{Palassini}%
  \BibitemOpen
  \bibfield  {author} {\bibinfo {author} {\bibfnamefont {M.}~\bibnamefont {Palassini}}\ and\ \bibinfo {author} {\bibfnamefont {A.}~\bibnamefont {Young}},\ }\bibfield  {title} {\bibinfo {title} {Triviality of the ground state structure in ising spin glasses},\ }\href@noop {} {\bibfield  {journal} {\bibinfo  {journal} {Phys. Rev. Lett.}\ }\textbf {\bibinfo {volume} {83}},\ \bibinfo {pages} {5126} (\bibinfo {year} {1999})}\BibitemShut {NoStop}%
\bibitem [{\citenamefont {Ebert}\ \emph {et~al.}(2022)\citenamefont {Ebert}, \citenamefont {Gessert},\ and\ \citenamefont {Weigel}}]{Ebert2022}%
  \BibitemOpen
  \bibfield  {author} {\bibinfo {author} {\bibfnamefont {P.}~\bibnamefont {Ebert}}, \bibinfo {author} {\bibfnamefont {D.}~\bibnamefont {Gessert}},\ and\ \bibinfo {author} {\bibfnamefont {M.}~\bibnamefont {Weigel}},\ }\bibfield  {title} {\bibinfo {title} {Weighted averages in population annealing: Analysis and general framework},\ }\href@noop {} {\bibfield  {journal} {\bibinfo  {journal} {Phys. Rev. E}\ }\textbf {\bibinfo {volume} {106}},\ \bibinfo {pages} {045303} (\bibinfo {year} {2022})}\BibitemShut {NoStop}%
\bibitem [{\citenamefont {Barzegar}\ \emph {et~al.}(2018)\citenamefont {Barzegar}, \citenamefont {Pattison}, \citenamefont {Wang},\ and\ \citenamefont {Katzgraber}}]{Barzegar2018}%
  \BibitemOpen
  \bibfield  {author} {\bibinfo {author} {\bibfnamefont {A.}~\bibnamefont {Barzegar}}, \bibinfo {author} {\bibfnamefont {C.}~\bibnamefont {Pattison}}, \bibinfo {author} {\bibfnamefont {W.}~\bibnamefont {Wang}},\ and\ \bibinfo {author} {\bibfnamefont {H.}~\bibnamefont {Katzgraber}},\ }\bibfield  {title} {\bibinfo {title} {Optimization of population annealing monte carlo for large-scale spin-glass simulations},\ }\href@noop {} {\bibfield  {journal} {\bibinfo  {journal} {Phys. Rev. E}\ }\textbf {\bibinfo {volume} {98}},\ \bibinfo {pages} {053308} (\bibinfo {year} {2018})}\BibitemShut {NoStop}%
\bibitem [{\citenamefont {Kramers}\ and\ \citenamefont {Wannier}(1941)}]{PhysRev.60.252}%
  \BibitemOpen
  \bibfield  {author} {\bibinfo {author} {\bibfnamefont {H.}~\bibnamefont {Kramers}}\ and\ \bibinfo {author} {\bibfnamefont {G.}~\bibnamefont {Wannier}},\ }\bibfield  {title} {\bibinfo {title} {Statistics of the two-dimensional ferromagnet. part i},\ }\href@noop {} {\bibfield  {journal} {\bibinfo  {journal} {Phys. Rev.}\ }\textbf {\bibinfo {volume} {60}},\ \bibinfo {pages} {252} (\bibinfo {year} {1941})}\BibitemShut {NoStop}%
\bibitem [{\citenamefont {Wegner}(1971)}]{Wegner:1971app}%
  \BibitemOpen
  \bibfield  {author} {\bibinfo {author} {\bibfnamefont {F.}~\bibnamefont {Wegner}},\ }\bibfield  {title} {\bibinfo {title} {Duality in generalized ising models and phase transitions without local order parameters},\ }\href@noop {} {\bibfield  {journal} {\bibinfo  {journal} {J. Math. Phys.}\ }\textbf {\bibinfo {volume} {12}},\ \bibinfo {pages} {2259} (\bibinfo {year} {1971})}\BibitemShut {NoStop}%
\bibitem [{\citenamefont {Wang}\ \emph {et~al.}(2003)\citenamefont {Wang}, \citenamefont {Harrington},\ and\ \citenamefont {Preskill}}]{WANG200331}%
  \BibitemOpen
  \bibfield  {author} {\bibinfo {author} {\bibfnamefont {C.}~\bibnamefont {Wang}}, \bibinfo {author} {\bibfnamefont {J.}~\bibnamefont {Harrington}},\ and\ \bibinfo {author} {\bibfnamefont {J.}~\bibnamefont {Preskill}},\ }\bibfield  {title} {\bibinfo {title} {Confinement-higgs transition in a disordered gauge theory and the accuracy threshold for quantum memory},\ }\href@noop {} {\bibfield  {journal} {\bibinfo  {journal} {Annals of Physics}\ }\textbf {\bibinfo {volume} {303}},\ \bibinfo {pages} {31} (\bibinfo {year} {2003})}\BibitemShut {NoStop}%
\bibitem [{\citenamefont {Ohno}\ \emph {et~al.}(2004)\citenamefont {Ohno}, \citenamefont {Arakawa}, \citenamefont {Ichinose},\ and\ \citenamefont {Matsui}}]{OHNO2004462}%
  \BibitemOpen
  \bibfield  {author} {\bibinfo {author} {\bibfnamefont {T.}~\bibnamefont {Ohno}}, \bibinfo {author} {\bibfnamefont {G.}~\bibnamefont {Arakawa}}, \bibinfo {author} {\bibfnamefont {I.}~\bibnamefont {Ichinose}},\ and\ \bibinfo {author} {\bibfnamefont {T.}~\bibnamefont {Matsui}},\ }\bibfield  {title} {\bibinfo {title} {Phase structure of the random-plaquette z2 gauge model: accuracy threshold for a toric quantum memory},\ }\href@noop {} {\bibfield  {journal} {\bibinfo  {journal} {Nuclear Physics B}\ }\textbf {\bibinfo {volume} {697}},\ \bibinfo {pages} {462} (\bibinfo {year} {2004})}\BibitemShut {NoStop}%
\bibitem [{\citenamefont {Takeda}\ and\ \citenamefont {Nishimori}(2004)}]{TAKEDA2004377}%
  \BibitemOpen
  \bibfield  {author} {\bibinfo {author} {\bibfnamefont {K.}~\bibnamefont {Takeda}}\ and\ \bibinfo {author} {\bibfnamefont {H.}~\bibnamefont {Nishimori}},\ }\bibfield  {title} {\bibinfo {title} {Self-dual random-plaquette gauge model and the quantum toric code},\ }\href@noop {} {\bibfield  {journal} {\bibinfo  {journal} {Nuclear Physics B}\ }\textbf {\bibinfo {volume} {686}},\ \bibinfo {pages} {377} (\bibinfo {year} {2004})}\BibitemShut {NoStop}%
\bibitem [{\citenamefont {Arakawa}\ and\ \citenamefont {Ichinose}(2004)}]{ARAKAWA2004152}%
  \BibitemOpen
  \bibfield  {author} {\bibinfo {author} {\bibfnamefont {G.}~\bibnamefont {Arakawa}}\ and\ \bibinfo {author} {\bibfnamefont {I.}~\bibnamefont {Ichinose}},\ }\bibfield  {title} {\bibinfo {title} {Zn gauge theories on a lattice and quantum memory},\ }\href@noop {} {\bibfield  {journal} {\bibinfo  {journal} {Annals of Physics}\ }\textbf {\bibinfo {volume} {311}},\ \bibinfo {pages} {152} (\bibinfo {year} {2004})}\BibitemShut {NoStop}%
\bibitem [{\citenamefont {Bortz}\ \emph {et~al.}(1975)\citenamefont {Bortz}, \citenamefont {Kalos},\ and\ \citenamefont {Lebowitz}}]{BORTZ1975}%
  \BibitemOpen
  \bibfield  {author} {\bibinfo {author} {\bibfnamefont {A.}~\bibnamefont {Bortz}}, \bibinfo {author} {\bibfnamefont {M.}~\bibnamefont {Kalos}},\ and\ \bibinfo {author} {\bibfnamefont {J.}~\bibnamefont {Lebowitz}},\ }\bibfield  {title} {\bibinfo {title} {A new algorithm for monte carlo simulation of ising spin systems},\ }\href@noop {} {\bibfield  {journal} {\bibinfo  {journal} {Journal of Computational Physics}\ }\textbf {\bibinfo {volume} {17}},\ \bibinfo {pages} {10} (\bibinfo {year} {1975})}\BibitemShut {NoStop}%
\bibitem [{\citenamefont {Baity-Jesi}\ \emph {et~al.}(2013)\citenamefont {Baity-Jesi}, \citenamefont {Ba\~nos}, \citenamefont {Cruz}, \citenamefont {Fernandez}, \citenamefont {Gil-Narvion}, \citenamefont {Gordillo-Guerrero}, \citenamefont {I\~niguez}, \citenamefont {Maiorano}, \citenamefont {Mantovani}, \citenamefont {Marinari}, \citenamefont {Martin-Mayor}, \citenamefont {Monforte-Garcia}, \citenamefont {Sudupe}, \citenamefont {Navarro}, \citenamefont {Parisi}, \citenamefont {Perez-Gaviro}, \citenamefont {Pivanti}, \citenamefont {Ricci-Tersenghi}, \citenamefont {Ruiz-Lorenzo}, \citenamefont {Schifano}, \citenamefont {Seoane}, \citenamefont {Tarancon}, \citenamefont {Tripiccione},\ and\ \citenamefont {Yllanes}}]{Baity_L40}%
  \BibitemOpen
  \bibfield  {author} {\bibinfo {author} {\bibfnamefont {M.}~\bibnamefont {Baity-Jesi}}, \bibinfo {author} {\bibfnamefont {R.~A.}\ \bibnamefont {Ba\~nos}}, \bibinfo {author} {\bibfnamefont {A.}~\bibnamefont {Cruz}}, \bibinfo {author} {\bibfnamefont {L.~A.}\ \bibnamefont {Fernandez}}, \bibinfo {author} {\bibfnamefont {J.~M.}\ \bibnamefont {Gil-Narvion}}, \bibinfo {author} {\bibfnamefont {A.}~\bibnamefont {Gordillo-Guerrero}}, \bibinfo {author} {\bibfnamefont {D.}~\bibnamefont {I\~niguez}}, \bibinfo {author} {\bibfnamefont {A.}~\bibnamefont {Maiorano}}, \bibinfo {author} {\bibfnamefont {F.}~\bibnamefont {Mantovani}}, \bibinfo {author} {\bibfnamefont {E.}~\bibnamefont {Marinari}}, \bibinfo {author} {\bibfnamefont {V.}~\bibnamefont {Martin-Mayor}}, \bibinfo {author} {\bibfnamefont {J.}~\bibnamefont {Monforte-Garcia}}, \bibinfo {author} {\bibfnamefont {A.~M.~n.}\ \bibnamefont {Sudupe}}, \bibinfo {author} {\bibfnamefont {D.}~\bibnamefont {Navarro}}, \bibinfo {author} {\bibfnamefont {G.}~\bibnamefont {Parisi}},
  \bibinfo {author} {\bibfnamefont {S.}~\bibnamefont {Perez-Gaviro}}, \bibinfo {author} {\bibfnamefont {M.}~\bibnamefont {Pivanti}}, \bibinfo {author} {\bibfnamefont {F.}~\bibnamefont {Ricci-Tersenghi}}, \bibinfo {author} {\bibfnamefont {J.~J.}\ \bibnamefont {Ruiz-Lorenzo}}, \bibinfo {author} {\bibfnamefont {S.~F.}\ \bibnamefont {Schifano}}, \bibinfo {author} {\bibfnamefont {B.}~\bibnamefont {Seoane}}, \bibinfo {author} {\bibfnamefont {A.}~\bibnamefont {Tarancon}}, \bibinfo {author} {\bibfnamefont {R.}~\bibnamefont {Tripiccione}},\ and\ \bibinfo {author} {\bibfnamefont {D.}~\bibnamefont {Yllanes}},\ }\bibfield  {title} {\bibinfo {title} {Critical parameters of the three-dimensional ising spin glass},\ }\href@noop {} {\bibfield  {journal} {\bibinfo  {journal} {Phys. Rev. B}\ }\textbf {\bibinfo {volume} {88}},\ \bibinfo {pages} {224416} (\bibinfo {year} {2013})}\BibitemShut {NoStop}%
\bibitem [{\citenamefont {Amey}\ and\ \citenamefont {Machta}(2018)}]{Amey2018}%
  \BibitemOpen
  \bibfield  {author} {\bibinfo {author} {\bibfnamefont {C.}~\bibnamefont {Amey}}\ and\ \bibinfo {author} {\bibfnamefont {J.}~\bibnamefont {Machta}},\ }\bibfield  {title} {\bibinfo {title} {Analysis and optimization of population annealing},\ }\href@noop {} {\bibfield  {journal} {\bibinfo  {journal} {Phys. Rev. E}\ }\textbf {\bibinfo {volume} {97}},\ \bibinfo {pages} {033301} (\bibinfo {year} {2018})}\BibitemShut {NoStop}%
\end{thebibliography}%

\end{document}